%% file: Tor_self_adjoint_v4_1.tex
\documentclass[a4paper,10pt]{article}
\usepackage[text={7in,9in},centering]{geometry}
\usepackage[utf8]{inputenc}
\input{color.tex}

\input{talk.tex}

\title{Observables compatible to the toroidal moment operator}
\author{Drago\c s-Victor Anghel\thanks{Institutul National de Cercetare-Dezvoltare pentru Fizica si Inginerie Nucleara Horia Hulubei, dragos@theory.nipne.ro} \ and Amanda Teodora Preda\thanks{Institutul National de Cercetare-Dezvoltare pentru Fizica si Inginerie Nucleara Horia Hulubei, University of Bucharest, Faculty of Physics, amanda.preda@theory.nipne.ro}}

\begin{document}

\maketitle

\begin{abstract}
The quantum operator $\hat{T}_3$, corresponding to the projection of the toroidal moment on the $z$ axis, admits several self-adjoint extensions, when defined on the whole $\R^3$ space.
$\hat{T}_3$ commutes with $\hat{L}_3$ (the projection of the angular momentum operator on the $z$ axis) and they have what we call a \textit{natural set of coordinates}, denoted $(k,u,\phi)$, where $\phi$ is the azimuthal angle.
The second set of \textit{natural coordinates} is $(k_1,k_2,u)$, where $k_1 = k\cos\phi$, $k_2 = k\sin\phi$.
In both sets these coordinates, the operators get the simple forms $\hat{T}_3 \equiv -i\hbar \partial/\partial u$ and $\hat{L}_3 = -i\hbar \partial/\partial\phi$.
In both sets, $\hat{T}_3 = -i\hbar\partial/\partial u$, so any operator that is a function of $k$ and the partial derivatives with respect to the \textit{natural variables} $(k, u, \phi)$ commute with $\hat{T}_3$ and $\hat{L}_3$.
Similarly, operators that are functions of $k_1$, $k_2$, and the partial derivatives with respect to $k_1$, $k_2$, and $u$ commute with $\hat{T}_3$.
Therefore, we introduce here the operators $\hat{p}_{k} \equiv -i \hbar \partial/\partial k$, $\hat{p}^{(k1)} \equiv -i \hbar \partial/\partial k_1$, and $\hat{p}^{(k2)} \equiv -i \hbar \partial/\partial k_2$ and express them in the $(x,y,z)$ coordinates.
One may also invert the relations and write the typical operators, like the momentum $\hat{\bf p} \equiv -i\hbar {\bf \nabla}$ or the kinetic energy $\hat{H}_0 \equiv \hbar^2\Delta/(2m)$ in terms of the ``toroidal'' operators $\hat{T}_3$, $\hat{p}^{(k)}$, $\hat{p}^{(k1)}$, $\hat{p}^{(k2)}$, and, eventually, $\hat{L}_3$.
The formalism may be applied to specific physical systems, like nuclei, condensed matter systems, or metamaterials.
We exemplify it by calculating the momentum operator and the free particle Hamiltonian in terms of \textit{natural coordinates} in a thin torus, where the general relations get considerably simplified.

\end{abstract}

\section{Introduction} \label{sec_intro}

\begin{figure}
	\centering
	\includegraphics[width=5 cm]{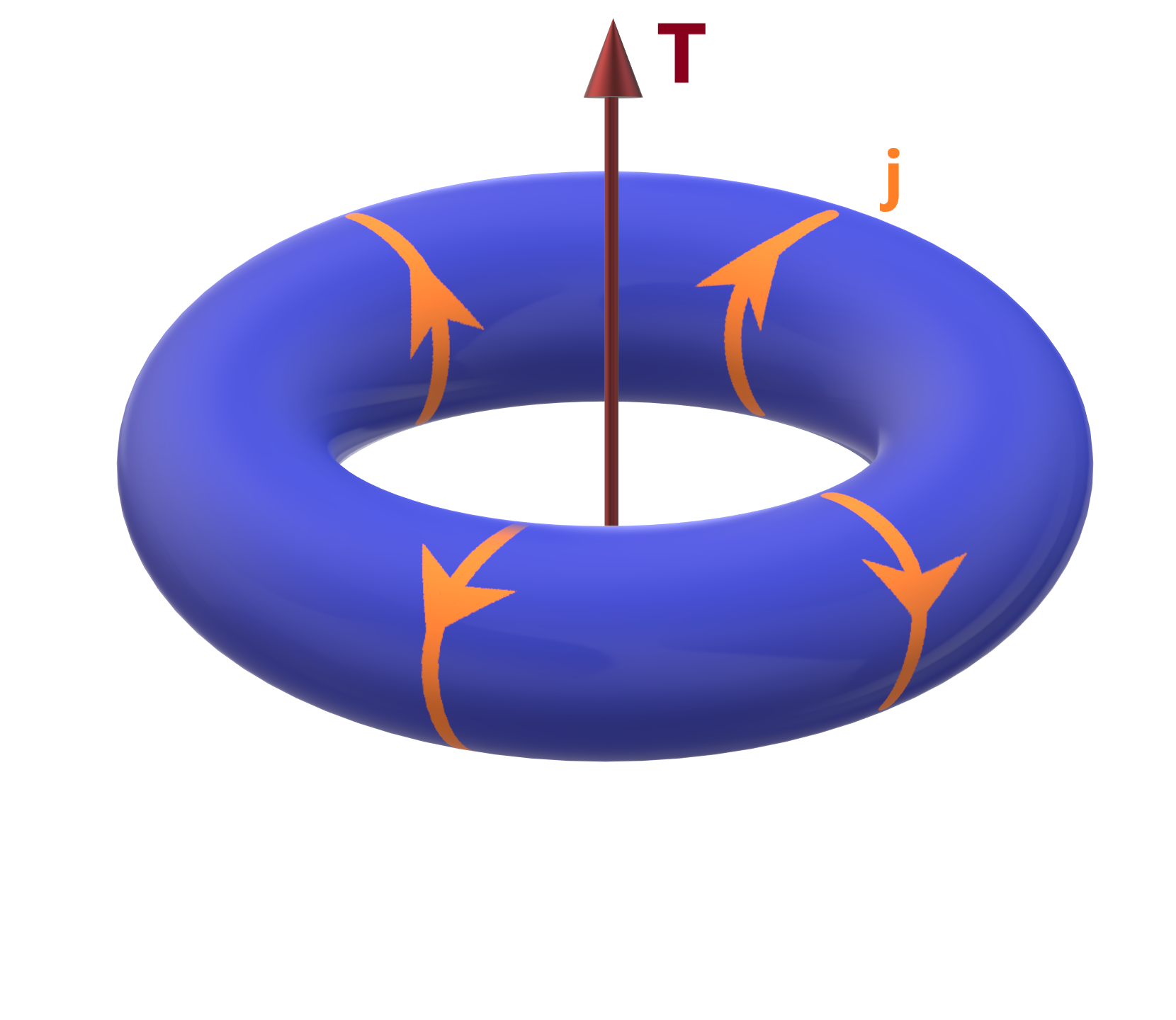}
    \caption{Toroidal dipole moment $\mathbf{T}$ generated by the currents $\mathbf{j}$ flowing on the surface of the torus}
	\label{fig_tor}
\end{figure}

In 1957, Zeldovich introduced a new type of electromagnetic interaction, in order to explain the parity nonconservation in $\beta$-decays~\cite{SovPhysJETP.6.1184.1958.Zeldovich}:
\begin{equation}
	\hat{H}_{\beta} \sim \bS \cdot \bJ^{ext} = \bS \cdot (\nabla \times H^{ext}) \label{H_Zeldovich}
\end{equation}
(where $\bS$ is the spin of the particle, $\bJ^{ext}$ is the external current, and $\bH^{ext}$ is the external magnetic field). Since neither electric nor magnetic multipoles lead to an interaction of the type~(\ref{H_Zeldovich}), he introduced for the first time the notion of "anapole", a type of distribution that he intuitively  described as a toroidal solenoid--a wire solenoid curved into a torus, as in Fig.~\ref{fig_tor}.
A magnetic field is produced inside this distribution, due to the static toroidal current, while the electric field everywhere is zero.

In the context of classical electrodynamics, Dubovik and Cheshkov proposed in the late 1960s that a whole new multipole family has to be introduced to complement the electric and magnetic moments ~\cite{Sov.24.1965.Dubovik, SovJ.5.318.1974.Dubovik}.
%
They reached the conclusion that a dynamic nonradiating anapole can only be achieved if toroidal moments are taken into consideration. Fundamental symmetry considerations explain the need for this new moment.
For example, the electric dipole moment is odd under spatial inversion and even under time reversal, while the magnetic dipole moment is odd under time reversal and even under spatial inversion. For a complete picture, one needs to include dipole moments which are even or odd under both transformations and these are the axial and polar toroidal dipole moments ~\cite{nanz:book.2016toroidal}.

Because it is difficult to isolate this weak radiative correction experimentally,  the anapole moment was measured in atomic Cesium by Wood \textit{et al} only decades after its theoretical prediction ~\cite{Science.275.1759.1997.Wood}. Although overlooked for a long time, it has been realized that the toroidal moments and anapole configurations emerged as an important tool to describe the properties of systems at all scales, from particle physics to the physics of macroscopic systems and this new field expanded rapidly ~\cite{NatureMat.15.263.2016.Papasimakis, AIPConfProc.477.14.1999.Flambaum}. In solid state physics, toroidal ordering was first studied theoretically by Charles Kittel ~\cite{PhysRev.70.965.1946.Kittel}  and it indicates the existence of a new type of magnetoelectric effect. In the framework of condensed matter, toroidal moments are linked to another kind of order parameter known as toroidization  (toroidal polarization). Media that exhibit macroscopic toroidization are called ferrotoroids and since they are expected to have promising technological applications (for example in data storage), it is crucial to investigate whether ferrotoroidicity is on an equal footing with the other feroic states (ferroelectric, ferromagnetic) ~\cite{Physics.2.20.2009.Khomskii, PU.55.557.2012.Pyatakov, JExpTPL.52.161.1990.Tolstoi, NewJPhys.9.95.2007.Fedotov, PhysRevB.84.094421.2011.Toledano, PhysRevB.101.2020.Shimada, JexpThPhys.87.146.1998.Popov, NatNano.14.141.2019.Lehmann}.

Recently, anapole states have generated tremendous interest in the field of nanophotonics, optics and metamaterials and have great potential for a wide range of applications like lasers, sensing and nonscattering objects that may be used for their cloaking behaviour ~\cite{Nanotechnology.30.2019.Yang, Comm.Pres.2.10.2019.Savinov, Nanophotonics.7.2017.Talebi, LasPhotRev.13.1800266.2019.Gurvitz}. It has even been proposed that since anapole states interact weakly with electomagnetic fields, they could be used to protect qubits from environmental disturbance ~\cite{ScRep.5.2016.Zagoskin}.

One of the most intriguing theoretical discoveries in this field was that particles that may be good candidates for dark matter constituents can only have toroidal moments ~\cite{PhysLettB.722.341.2013.Ho, NuclPhysB.907.1.2016.Cabral}. Also, the CPT invariance requires that the electromagnetic structure of Majorana fermions consists only of toroidal moments~\cite{PhysRevD.32.1266.Radescu,ModPhysA.13.5257.1998.Dubovik}.

If the toroidal moments correspond to a quantum observable, the operator should be self-adjoint. A quantum operator corresponding to the projection of the toroidal moment on the $z$ axis was introduced in ~\cite{AnnPhys.209.13.1991.Costescu}, where the toroidal polarizability of H-like atoms was also calculated. In ~\cite{JPA30.3515.1997.Anghel} it was shown that the operator is hypermaximal (it admits several self-adjoint extensions) and a complete set of eigenfunctions was found.

The lowest order toroidal moment corresponding to a current distribution $\bj(\br)$ is a polar vector of components
\begin{equation} \label{def_Ti}
    T_i = \frac{1}{10 c} \int_V [r_i (\br \bj) - 2 r^2 j_i] d^3\br
\end{equation}
where $r \equiv |\br|$, $j = |\bj|$, $T \equiv |\bT|$, and $i=1,2,3$ number the components of a vector on the three axes $x$, $y$, and $z$: $\bT \equiv (T_1, T_2, T_3)$, $\bj \equiv (j_1, j_2, j_3)$, $\br \equiv (r_1, r_2, r_3) \equiv (x,y,z)$ ~\cite{PhysRep.187.145.1990.Dubovik}.
The electromagnetic interaction Hamiltonian
\begin{equation}
    \cH \equiv \int \left(\rho \phi - \frac{1}{c} \bj \bA \right) d^3 \br \label{def_H_em}
\end{equation}
may be decomposed into multipoles interactions, leading to the term characterizing the interaction of the toroidal moment with the external fields ~\cite{PhysRep.187.145.1990.Dubovik}
\begin{equation}
    \cH_{tor}(t) = -\bT(t) \left[\nabla \times \nabla \times A^{ext}(\br, t)\right]_{r=0}
    \equiv - \bT(t) \left[ \frac{4\pi}{c} \bJ^{ext}(\br, t) + \frac{1}{c} \dot{\bD}^{ext}(\br, t) \right]_{r=0} .
\end{equation}

To correctly predict the toroidal moments of quantum systems (particles, nuclei, or macroscopic systems), one has first to study the quantum operators corresponding to them and their basic mathematical properties.
%
The quantum operators which correspond to the classical toroial moments (\ref{def_Ti}) are~\cite{AnnPhys.209.13.1991.Costescu, JPA30.3515.1997.Anghel}
\begin{equation} \label{def_Ti_op}
\hat T_i \equiv \frac{1}{10 mc} \sum_{j=1}^{3} \left(x_i x_j - 2 r^2 \delta_{ij} \right) \hat p_j ,
\end{equation}
where $\hat p_j \equiv - i\hbar \partial/\partial x_j$ is the momentum operator along the $j$ axis and $m$ is the mass of the particle.
We observe that the commutation relations~\cite{JPA30.3515.1997.Anghel}
\begin{equation} \label{commut_TiLj}
[\hat T_i, \hat L_j] = i \hbar \epsilon_{ijk} \hat T_k ,
\end{equation}
are satisfied, where $\hat{L}_j$ is the projection of the angular momentum operator on the $j$ axis.
In Ref.~\cite{JPA30.3515.1997.Anghel} it was shown that in $\R^3$ space, $\hat T_3$ (and, equivalently, $\hat T_1$ and $\hat T_2$) is a hypermaximal operator (admits several self-adjoint extensions) and may correspond to a physical quantity.

The operator $\hat{T}_3$ has a specific ``\textit{natural coordinate}'', denoted by $u$, such that 
\begin{equation} \label{def_T3_op_nat}
\hat T_3 = -i\hbar \frac{\partial}{\partial u} ,
\end{equation}
(notice that in~\cite{JPA30.3515.1997.Anghel} $u$ had a different scale, namely $\hat{T}_3 = - [i\hbar / (10mc)] \partial / \partial u$).
Starting from $u$, one can form the so called ``\textit{natural systems of coordinates}'' for the operator $\hat{T}_3$, like $(k,u,\phi)$ and $(k_1, k_2, u)$, as it will be explained in Section~\ref{sec_method}.
In the \textit{natural systems of coordinates} it is easy to construct operators that commute with $\hat{T}_3$, so one can form complete systems of commuting observables (CSCO) and find their eigenvectors and eigenvalues.
Furthermore, we can translate the operators defined in the \textit{natural system of coordinates} into operators acting on the $(x,y,z)$ coordinates and vice-versa.

In this paper, we work in both sets of coordinates $(k,u,\phi)$ and $(k_1, k_2, u)$ and we express the partial derivatives in these coordinates as operators in the $(x,y,z)$ space.
Vice-versa, we express the $(x,yz)$ and $(\rho,u,\phi)$ coordinates and partial derivatives with respect to them, in terms of the $(k,u,\phi)$ and $(k_1, k_2, u)$ coordinates, respectively, and partial derivatives with respect to them.
This enables us to express the momentum operator and the free particle Hamiltonian as operators in \textit{natural coordinates} and analyze systems of practical importance.

The paper is organized as follows.
In Section~\ref{sec_method} we present the methods employed.
In Section~\ref{subsec_nat_var}, we present the natural sets of coordinates of the operator $\hat T_3$, denoted $(k, u, \phi)$ and $(k_1, k_2, u)$, in which $\hat T_3 = -i\hbar \partial/\partial u$ ($\phi$ is the azimuthal angle), whereas in Sections~\ref{subsec_Hilbert_sp} and \ref{subsec_self_ad} we analyze the Hilbert spaces in which we work  and the self-adjointness of $\hat{T}_3$.
In Section~\ref{sec_results} we introduce the new operators $\hat{p}^{(k)}$, $\hat{p}^{(k1)}$, and $\hat{p}^{(k2)}$, that commute with $\hat{T}_3$ ($\hat{p}^{(k)}$ commutes with both, $\hat{T}_3$ and $\hat{L}_3$).
These operators were found in the natural coordinates of $\hat{T}_3$ and subsequently they were expressed in the $(\rho,z,\phi)$ cylindrical coordinates or in the $(x,y,z)$ Cartesian coordinates.
The relations are then inverted, to express the familiar operators, like the momentum operator and the Hamiltonian, in terms of operators in the \textit{natural coordinates}.
The results are discussed in Section~\ref{sec_discussion}, where we apply the formalism to a thin torus, whereas in Section~\ref{sec_conclusions} are given the conclusions.
In the Appendix are shown the calculations details.

\section{Method} \label{sec_method}

The main tools used in this paper, as in Ref.~\cite{JPA30.3515.1997.Anghel}, are two \textit{natural sets of coordinates} of the toroidal moment operator, which we shall present in this section to introduce the notations and make the paper easier to read.
Then, we present the Hilbert spaces on which we apply the operators we introduce in the most general case, to analyze if they are essentially self-adjoint and if they may correspond to observables.

\subsection{The natural coordinates for the toroidal moment operator} \label{subsec_nat_var}

\begin{figure}
    \centering
    \includegraphics[width=5 cm]{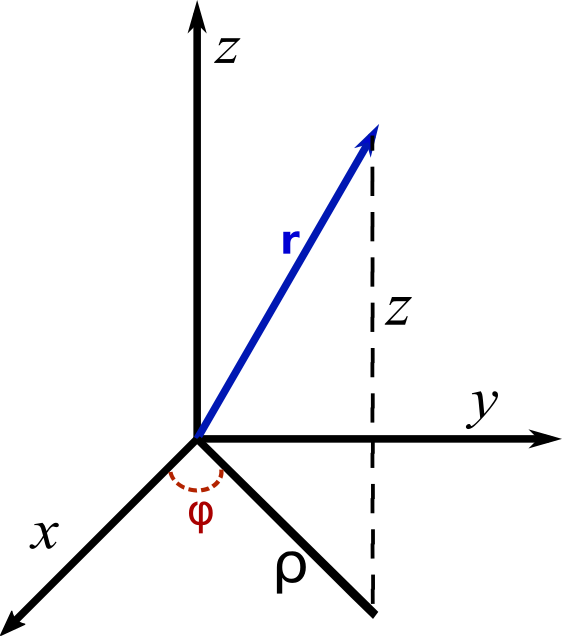}
    \caption{The notations of the Cartesian and cylindrical systems of coordinates.}
    \label{fig_cyl_coord}
\end{figure}

\begin{figure}
    \centering
    \includegraphics[width=7cm, keepaspectratio=true]{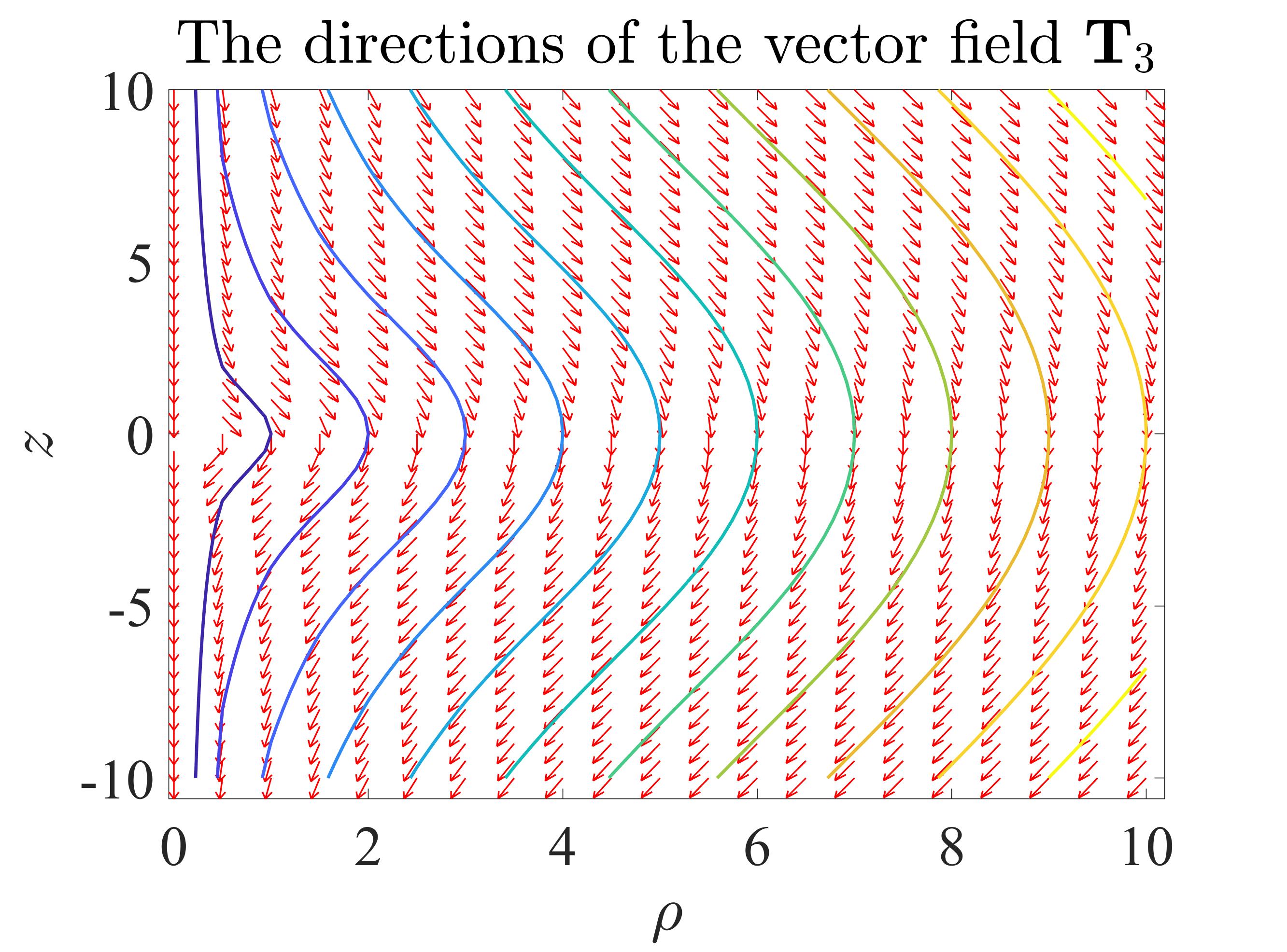}
    \includegraphics[width=7cm, keepaspectratio=true]{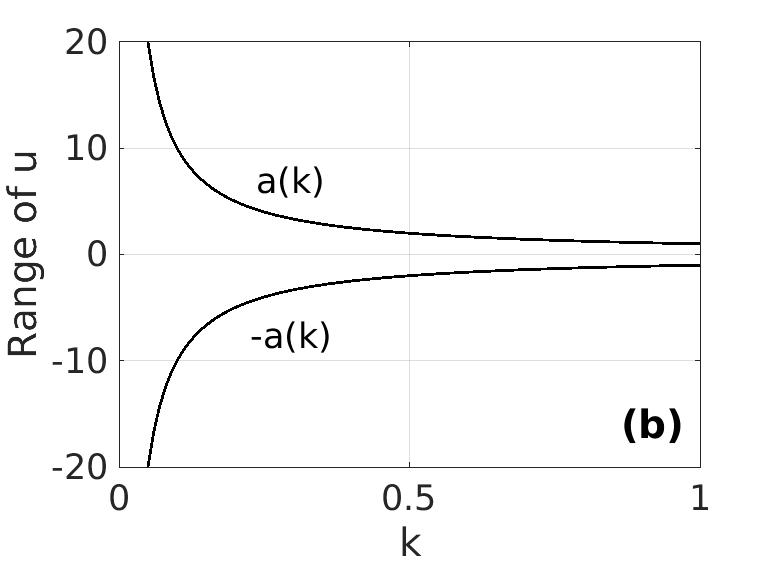}
    \caption{(a) The directions of the vector field $\bT_3$ (i.e., $\bt_3 \equiv \bT_3/T_3$) in a plane that contains the $z$ axis (the units on the axes are arbitrary).
        The vector field is tangent in every point to the continuous lines that represent the lines along which the ``natural variable'' $u$ vary.
        In (b) we show the range of $u$, as a function of $k$ (in arbitrary units): $u\in(-a(k), a(k))$.
    }
    \label{fig_coordinates}
\end{figure}

In the following we shall focus on the operator $\hat T_3$.
The analysis of $\hat{T}_1$ and $\hat{T}_2$ may be done in a similar fashion.
According to Eq.~(\ref{commut_TiLj}), $[\hat{T}_3, \hat{L}_3] = 0$, so it is more convenient to write in cylindrical coordinates~\cite{JPA30.3515.1997.Anghel}
\begin{equation} \label{def_T3_op_cyl}
\hat T_3 = \frac{-i\hbar}{10 mc} \left[ z\rho \frac{\partial}{\partial \rho} - \left( 2\rho^2 + z^2 \right) \frac{\partial}{\partial z} \right] ,
\end{equation}
where $\rho \equiv \sqrt{x^2 + y^2}$, $r = \sqrt{\rho^2 + z^2}$ (see Fig.~\ref{fig_cyl_coord}).
We define the vector field $\bT_3$ by the relation
\begin{equation}
    \hat{T}_3 \equiv \bT_3 \cdot \nabla . \label{def_bTi}
\end{equation}
We plot the directions of $\bT_3$ (i.e. $\bt_3 \equiv \bT_3/T_3$, where $T_3 \equiv |\bT_3|$) in Fig.~\ref{fig_coordinates}, in a plane which contains the $z$ axis.
We observe that $\hat{T}_3$ may be written as a derivative along the lines of this  vector field, that is, along the continuous curves shown in the figure.
This observation led to the introduction of the ``natural'' set of coordinates $(k,u,\phi)$~\cite{JPA30.3515.1997.Anghel}, in which $\hat{T}_3$ takes the form~(\ref{def_T3_op_nat}), whereas
%
\begin{equation}
    k \equiv \left[ \rho^2 \left( z^2 + \rho^2 \right) \right]^{1/4} \equiv \sqrt{\rho r} \quad {\rm and} \quad
    u \equiv - 10 mc \int_0^z \frac{dt}{\sqrt{t^4 + 4 k^4}} = \pm 10 mc \int_\rho^k \frac{dt}{\sqrt{-t^4 + k^4}} ; \label{def_nat_set}
\end{equation}
the curves in Fig.~\ref{fig_coordinates} correspond to $k(\rho,z) = {\rm const}$.

The variable $u$ takes values in a finite interval $(-a(k), a(k))$, with~\cite{JPA30.3515.1997.Anghel}
\begin{equation} \label{def_ak}
    a(k) = \frac{10 mc}{4k} \frac{\Gamma\left(\frac{1}{4}\right) \Gamma\left(\frac{1}{2}\right)}{\Gamma\left(\frac{3}{4}\right)}
    \equiv \frac{10 mc}{k} C_a
\end{equation}
and $C_a \approx 1.31103$.
Notice that all the end points $-a(k)$ and $a(k)$ (for all $k$) correspond to the points $(z\to \infty, \rho = 0)$ and $(z\to -\infty, \rho = 0)$, respectively.

Beside $(k,u,\phi)$, we may define the set of coordinates $(k_1, k_2, u)$ by
\begin{equation}
    k_1 = k\cos\phi \qquad {\rm and} \qquad k_2 = k\sin\phi, \label{def_k1k2}
\end{equation}
in which the relation~(\ref{def_T3_op_nat}) is still valid.
In these variables, $k_1$ and  $k_2$ take values in the interval $(-\infty, \infty)$, whereas $u$ takes values in the interval $(-a(k_1^2+k_2^2), a(k_1^2+k_2^2))$ (Eq.~\ref{def_ak}).

\subsection{The Hilbert space} \label{subsec_Hilbert_sp}

On the space of integrable single valued complex functions defined on $\R^3$, we define the scalar product of two functions $f$ and $g$ by
\begin{equation} \label{def_sc_prod}
    \langle f | g \rangle \equiv \int_{\R^3} f^*(\br) g(\br) d^3 \br
\end{equation}
and the norm
\begin{equation} \label{def_norm}
|| f || \equiv \sqrt{\langle f | f \rangle} .
\end{equation}
We work with the Hilbert space $\hH \equiv L^2(\R^3)$, which consists of the functions integrable in modulus square.

The space $\R^3$ minus the $z$ axis, $\tilde{\R}^3 \equiv \R^3\backslash z$, is mapped onto a set $\mathbb{M}$ in the $(k,u,\phi)$ space, where $\phi \in [0, 2\pi)$, $k\in(0,\infty)$, and $a\in (-a(k), a(k))$.
Then, the functions $f(\br),\ g(\br)$, from $\hH$, are mapped into the functions $\tilde{f}(k,u,\phi),\ \tilde{g}(k,u,\phi)$ defined on the set $\mathbb{M}$.
The volume element is transformed as~\cite{JPA30.3515.1997.Anghel}
\begin{equation} \label{vol_elem}
dx dy dz = \frac{k^3}{5 m c} dk du d\phi \equiv I_1 dk du d\phi .
\end{equation}
so the scalar product and the norm (Eqs~\ref{vol_elem} and \ref{def_sc_prod}) are
%
\begin{equation} \label{def_sc_prod_ukf}
\langle f | g \rangle = \langle \tilde{f} | \tilde{g} \rangle = \int\limits_0^{2\pi} d\phi \int\limits_0^\infty dk\, \frac{k^3}{5 mc} \int\limits_{-a(k)}^{a(k)} du\, \tilde{f}^*(u,k,\phi) \tilde{g}(u,k,\phi)
\quad {\rm and} \quad
||\tilde{f}|| \equiv \sqrt{\langle\tilde{f} | \tilde{f}\rangle} = ||f||,
\end{equation}
respectively.
Then, by $\tilde{\hH} \equiv L^2(\mathbb{M})$ we denote the Hilbert space of functions of finite norms defined on $\mathbb{M}$, which is the image of $\hH(\R^3)$.

In the coordinates $(k_1,k_2, u)$~(\ref{def_k1k2}) we have the relation
\begin{equation} \label{vol_elem2}
dx dy dz = \frac{2k^2}{10 mc} dk_1 dk_2 d\phi \equiv I_2 dk_1 dk_2 d\phi ,
\end{equation}
together with the scalar product and the norm,
\begin{equation} \label{def_sc_prod_uk1k2}
\langle f | g \rangle = \langle \tilde{\tilde{f}} | \tilde{\tilde{g}} \rangle
= \int\limits_{-\infty}^{\infty} dk_1 \int\limits_{-\infty}^{\infty} dk_2 \, (k_1^2 + k_2^2) \int\limits_{-a(\sqrt{k_1^2 + k_2^2})}^{a(\sqrt{k_1^2 + k_2^2})} du \, \tilde{\tilde{f}}^*(k_1,k_2,u) \tilde{\tilde{g}}(k_1,k_2,u)
\quad {\rm and} \quad
||\tilde{\tilde{f}}|| \equiv \sqrt{\langle\tilde{\tilde{f}} | \tilde{\tilde{f}}\rangle},
\end{equation}
respectively.
The domain of definition, $k_1, k_2 \in (-\infty,\infty)$ and $a\in\left( -a(\sqrt{k_1^2+k_2^2}), a(\sqrt{k_1^2+k_2^2}) \right)$, will be denoted by $\M_1$ and the Hilbert space of functions of finite norms on $\M_1$ is denoted by $\tilde{\tilde{\cH}} = L^2(\M_1)$.

\subsection{Self-adjointness and eigenfunctions of $\hat T_3$} \label{subsec_self_ad}

Any observable should correspond to a self-adjoint operator in quantum mechanics.
In both sets of coordinates, $(k,u,\phi)$ and $(k_1,k_2,u)$, $\hat T_3 = - i\hbar \partial/\partial u$.
The sets $\M$ (in $(k,u,\phi)$) and $\M_1$ (in $(k_1,k_2,u)$) correspond to the whole $\R^3$ $(x,y,z)$ space and no boundary conditions are imposed on their frontiers--that is, when $u = \pm a(k)$ in $(k,u,\phi)$ and $u = \pm a(\sqrt{k_1^2+k_2^2})$ in $(k_1,k_2,u)$.
Then, in the coordinates $(k,u,\phi)$, $\hat T_3$ si defined on the set of smooth functions on $\M$ and the operator is essentially self-adjoint, since we can impose the boundary conditions~\cite{JPA30.3515.1997.Anghel}
\begin{subequations} \label{b_cond_MM1}
\begin{equation}
    \tilde{f}[k, - a(k), \phi] = \tilde{f}[k, a(k), \phi] e^{i\theta(k,\phi)} , \label{b_cond_M}
\end{equation}
where $\theta(k,\phi)$ can be any function of $k$ and $u$.
Of course, the same conclusion is reached working in the $(k_1,k_2,u)$ space, where the boundary conditions read
\begin{equation}
\tilde{\tilde{f}} \left[k_1,k_2, - a(\sqrt{k_1^2+k_2^2}) \right] = \tilde{\tilde{f}} \left[k_1,k_2, a(\sqrt{k_1^2+k_2^2}) \right] e^{i\theta(k,\phi)} . \label{b_cond_M1}
\end{equation}
\end{subequations}
%
A set of orthonormal eigenfunctions for the $\hat{T}_3$ operator in the $(k,u,\phi)$ space, which satisfy periodic boundary conditions in the $u$ direction, was proposed in Ref.~\cite{JPA30.3515.1997.Anghel}:
\begin{equation}
\cT_{k_0, t_{k_0}, m} (k,u,\phi) \equiv \frac{1}{2 k \sqrt{2 \pi C_a}} \delta(k-k_0) e^{\frac{i}{\hbar} t_{k_0} u} e^{\frac{i}{\hbar} m\phi} , \label{def_eigenf}
\end{equation}
where $k_0 \in (0, \infty)$, $t_{k_0} \equiv n \pi \hbar/a(k_0)$, whereas $m$ and $n$ are integers.
We see that the functions~(\ref{def_eigenf}) are localized on the $k$ axis of the space $\M$.

If $\hat{T}_3$ is not defined on the whole space $\R^3$, then different boundary conditions may be imposed.
For example, the definition domain may be a finite region $\D$ of $\R^3$, with Dirichlet boundary conditions on the frontier (e.g. a nucleus, a region of a condensed matter system, or an element of a metamaterial).

\section{Results} \label{sec_results}

In order to find other operators that commute with $\hat{T}_3$ (beside $\hat{L}_3$ and the ``position'' operator $\hat{k}$), we notice that in the space $(k,u,\phi)$ any operator which is a function of the operators $k$, $\partial / \partial k$, $\partial/\partial u$, and $\partial/\partial \phi$ commutes with both, $\hat{T}_3$ and $\hat{L}_3$.
Similarly, in the space $(k_1, k_2, u)$, any operator that is a function of $k_1$, $k_2$, $\partial/\partial k_1$,  $\partial/\partial k_2$, and $\partial/\partial u$ commutes with $\hat{T}_3$.
In this section, we calculate the operators $\partial/\partial k$ (in the $(k,u,\phi)$ space), $\partial/\partial k_1$, and $\partial/\partial k_2$ (in the $(k_1, k_2,u)$ space).
Having these, we can form complete sets of commuting observables involving partial derivatives, of eigenvectors which are not localized in space, like the ones given in~(\ref{def_eigenf}).
For example, in the coordinates $(k_1, k_2,u)$, the operators $\hat{p}^{(k1)} \equiv -i\hbar \partial / \partial k_1$, $\hat{p}^{(k2)} \equiv -i\hbar \partial / \partial k_2$, and $\hat{T}_3 \equiv -i\hbar \partial / \partial u$ resemble the three components of a ``momentum operator'' 
and form a CSCO.
On the other hand, in the coordinates $(k,u,\phi)$, the operators $\hat{T}_3 \equiv -i\hbar \partial/\partial u$ and $\hat{L}_3 \equiv -i\hbar \partial / \partial \phi$ correspond to two observables (the projections on the $z$ axis of the toroidal moment and of the angular momentum), whereas the third operator, defined as $\hat{p}^{(k)} \equiv -i\hbar \partial / \partial k$, represents the projection of the ``momentum operator'' in the $(k_1,k_2,u)$ coordinates, on the radial direction $k \equiv \sqrt{k_1^2+k_2^2}$ and is not an observable because it is not self-adjoint and it does not have the form of a ``radial momentum''.


\subsection{The operator $\hat p^{(k)}$, in the $(k,u,\phi)$ space} \label{subsec_hatK}

\begin{figure}
    \centering
    \includegraphics[width=7cm]{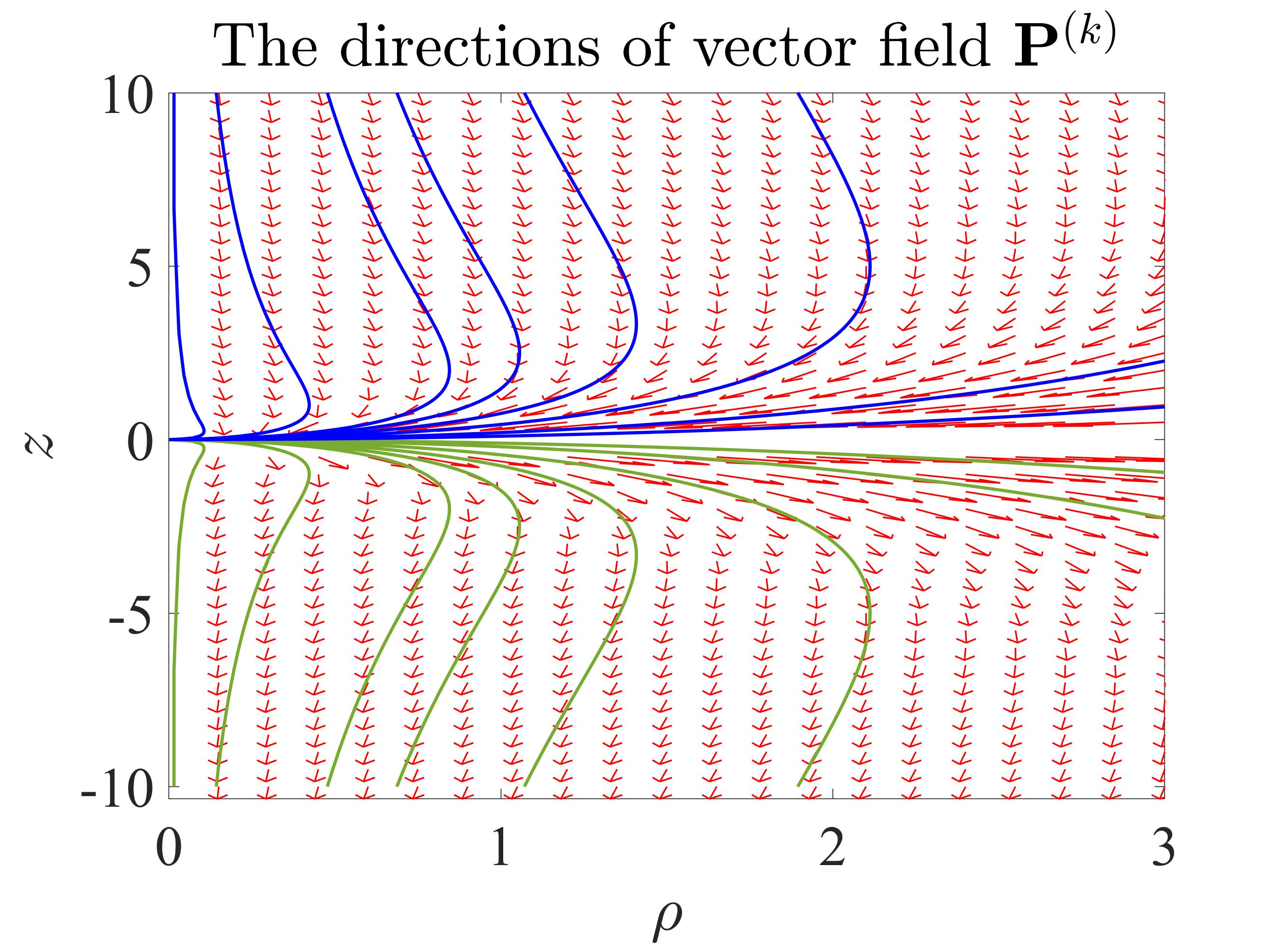}
    \caption{The directions of the vector field $\bP^{(k)}$ (i.e., $\bp^{(k)} \equiv \bP^{(k)}/P^{(k)}$) and the curves of constant $u(\rho, z)$.
    The vector field is tangent to the curves in every point.}
    \label{fig_Pk}
\end{figure}

The operator $\hat{p}^{(k)} = - i \hbar \partial / \partial k$ in the $(k,u,\phi)$ space may be expressed in terms of the $(x,y,z)$ coordinates and has the form (see Appendix~\ref{app_pk} for details)
\begin{equation}
    \hat{p}^{(k)} = - i \hbar \frac{\partial}{\partial k}
    = - i\hbar
    \frac{\partial f_u}{\partial k} \left(\frac{\partial f_u}{\partial z}\right)^{-1}
    \left[ \frac{(\partial f_u / \partial z) (\partial f_u/\partial k)^{-1} + \partial k/\partial z}{\partial k/ \partial \rho} \frac{\partial}{\partial \rho} - \frac{\partial}{\partial z} \right]
    \equiv \bP^{(k)} \cdot \hat{\bp}
    , \label{def_pk}
\end{equation}
where $\hat{\bp} \equiv - i\hbar \nabla$ is the momentum operator (in the $(x,y,z)$ coordinates) and the partial derivatives that appear in the expression above are given in Eqs.~(\ref{part_derivs_uk}).
The vector field $\bP^{(k)}$ in cylindrical coordinates is
\begin{equation}
    \bP^{(k)} \equiv \frac{\partial f_u}{\partial k} \left(\frac{\partial f_u}{\partial z}\right)^{-1}
    \left[ \frac{(\partial f_u / \partial z) (\partial f_u/\partial k)^{-1} + \partial k/\partial z}{\partial k/ \partial \rho} \hat{\bf\rho} - \hat{\bz} \right] .
    \label{def_Pk}
\end{equation}
%
The directions of $\bP^{(k)}$ (that is, $\bp^{(k)} \equiv \bP^{(k)}/P^{(k)}$) are drawn in Fig.~\ref{fig_Pk}, together with the curves of constant $u(\rho,z)$.
These curves correspond to lines parallel to the $k$ axis in the $(k,u,\phi)$ space (fixed $u$ and $\phi$), so the vector field $\bP^{(k)}$ is tangent to them in every point--similarly to $\bT_3$ in Fig.~\ref{fig_coordinates}.

Equations~(\ref{def_T3_op_cyl}) and (\ref{def_pk}) may be joined into a system from which we can calculate the partial derivatives in the $(x,y,z)$ coordinates in terms of the partial derivatives in the $(k,u,\phi)$ coordinates, by using Eqs.~(\ref{rho_ku}) and (\ref{derivs_zrho_k}).
The system of equations thus formed is
%
%
\begin{eqnarray}
\left[ \begin{array}{c}  \frac{\partial}{\partial k} \\ \frac{\partial}{\partial u} \end{array}\right]
&\equiv&
\left[\left[ T_3^{(k)} \right]\right]
\left[\begin{array}{c} \frac{\partial}{\partial \rho} \\ \frac{\partial}{\partial z} \end{array}\right]
\label{sys_part_cyl}
\end{eqnarray}
where by $[[\cdot]]$ we denote a square matrix and
\begin{eqnarray}
    \left[\left[ T_3^{(k)} \right]\right] \equiv \left[
    \begin{array}{cc}
    \frac{\partial \rho(k,u)}{\partial k} & \frac{\partial z(k,u)}{\partial k} \\
    \frac{z(k,u) \rho(k,u)}{10mc} & - \frac{2\rho^2+z^2}{10mc}
    \end{array}
    \right]
\end{eqnarray}
By inverting the system~(\ref{sys_part_cyl}), we write $\partial/\partial\rho$ and $\partial/\partial z$ in terms of $\partial/\partial k$ and $\partial/\partial u$, in the natural set of coordinates for $\hat{T}_3$.



\subsection{The operators $\hat p^{(k1)}$ and $\hat p^{(k2)}$, in the $(k_1,k_2,u)$ space} \label{subsec_hatK1K2}

If we switch from the coordinates $(k,u,\phi)$ to the coordinates $(k_1,k_2,u)$  we have
%
\begin{equation}
u \equiv - 10 mc \int_0^z \frac{dt}{\sqrt{t^4 + 4 (k_1^2 + k_2^2)^2}}
\equiv f_{u2} (z, k_1, k_2)
\quad {\rm and} \quad
a(k) = \frac{10 mc}{\sqrt{k_1^2 + k_2^2}} C_a .
\label{def_nat_set_k1k2}
\end{equation}
The coordinates $k_1 = k \cos\phi$ and $k_2 = k \sin\phi$ in terms of $x$, $y$, and $z$ are
\begin{equation} \label{defs_k1k2}
k_1
= x \left[ 1 + \frac{ z^2 }{x^2 + y^2} \right]^{1/4}
\equiv x \sqrt{\frac{r}{\rho}}
\qquad {\rm and} \qquad
k_2
= y \left[ 1 + \frac{ z^2 }{x^2 + y^2} \right]^{1/4}
\equiv y \sqrt{\frac{r}{\rho}} .
\end{equation}
%
Hence, we define the operators
\begin{equation}
    \hat p^{(k1)} \equiv -i\hbar \frac{\partial}{\partial k_1}
    \qquad {\rm and} \qquad
    \hat p^{(k2)} \equiv -i\hbar \frac{\partial}{\partial k_2} ,
    \label{def_pk1pk2}
\end{equation}
which, by definition, satisfy the commutation relations
\begin{equation}
    [\hat{T}_3, \hat{p}^{(k1)}] = [\hat{T}_3, \hat{p}^{(k2)}] = [\hat{p}^{(k1)}, \hat{p}^{(k2)}] = 0
    \label{comms_T3pk1pk2}
\end{equation}
The operators $\hat{p}^{(k1)}$ and $\hat{p}^{(k2)}$ are calculated in the Appendix~\ref{app_pk1pk2} and in the $(x,y,z)$ coordinates have the expressions
\begin{subequations} \label{pk1_pk2_xyz}
\begin{eqnarray}
    \hat{p}^{(k1)} &\equiv& - i\hbar \frac{\partial}{\partial k_1}
    = - i\hbar
    \frac{\partial f_{u2}}{\partial k_1} \left(\frac{\partial f_{u2}}{\partial z}\right)^{-1}
    \left\{
    \frac{\frac{\partial k_1}{\partial y} \frac{\partial k_2}{\partial z} \left(\frac{\partial k_2}{\partial y}\right)^{-1} - \frac{\partial f_{u2}}{\partial z} \left(\frac{\partial f_{u2}}{\partial k_1}\right)^{-1} - \frac{\partial k_1}{\partial z}}{\frac{\partial k_1}{\partial x} - \frac{\partial k_1}{\partial y} \frac{\partial k_2}{\partial x} \left(\frac{\partial k_2}{\partial y}\right)^{-1}}
    \frac{\partial}{\partial x} \right. \nonumber \\
    && \left. - \left(\frac{\partial k_2}{\partial y}\right)^{-1} \left[ \frac{\partial k_2}{\partial x} \frac{\frac{\partial k_1}{\partial y} \frac{\partial k_2}{\partial z} \left(\frac{\partial k_2}{\partial y}\right)^{-1} - \frac{\partial f_{u2}}{\partial z} \left(\frac{\partial f_{u2}}{\partial k_1}\right)^{-1} - \frac{\partial k_1}{\partial z}}{\frac{\partial k_1}{\partial x} - \frac{\partial k_1}{\partial y} \frac{\partial k_2}{\partial x} \left(\frac{\partial k_2}{\partial y}\right)^{-1}} + \frac{\partial k_2}{\partial z} \right]
    \frac{\partial}{\partial y}
    + \frac{\partial}{\partial z}
    \right\}
    \equiv \bP^{(k1)} \cdot \hat{\bp} ,
     \label{pk1_xyz} \\
    \hat{p}^{(k2)} &=& - i\hbar
    \frac{\partial f_{u2}}{\partial k_2} \left(\frac{\partial f_{u2}}{\partial z}\right)^{-1}
    \left\{
    - \frac{\frac{\partial k_2}{\partial y} \frac{\partial k_1}{\partial z} \left(\frac{\partial k_1}{\partial y}\right)^{-1} - \frac{\partial f_{u2}}{\partial z} \left(\frac{\partial f_{u2}}{\partial k_2}\right)^{-1} - \frac{\partial k_2}{\partial z}}
    {\frac{\partial k_2}{\partial x} - \frac{\partial k_2}{\partial y} \frac{\partial k_1}{\partial x} \left(\frac{\partial k_1}{\partial y}\right)^{-1}}
    \frac{\partial}{\partial x} \right. \nonumber \\
    && \left. + \left(\frac{\partial k_1}{\partial y}\right)^{-1} \left[ \frac{\partial k_1}{\partial x}
    \frac{\frac{\partial k_2}{\partial y} \frac{\partial k_1}{\partial z} \left(\frac{\partial k_1}{\partial y}\right)^{-1} - \frac{\partial f_{u2}}{\partial z} \left(\frac{\partial f_{u2}}{\partial k_2}\right)^{-1} - \frac{\partial k_2}{\partial z}}
    {\frac{\partial k_2}{\partial x} - \frac{\partial k_2}{\partial y} \frac{\partial k_1}{\partial x} \left(\frac{\partial k_1}{\partial y}\right)^{-1}}
    + \frac{\partial k_1}{\partial z} \right]
    \frac{\partial}{\partial y}
    - \frac{\partial}{\partial z}
    \right\}
    \equiv \bP^{(k2)} \cdot \hat{\bp}
     , \label{pk2_xyz}
\end{eqnarray}
\end{subequations}
where 
the partial derivatives are given in Eqs.~(\ref{part_derivs}).

The operators $(\hat{p}^{(k1)}, \hat{p}^{(k2)}, \hat{T}_3)$ form a CSCO.
In the $(k_1, k_2, u)$ space, the eigenfunctions corresponding to the $\hat{p}^{(k1)}$, $\hat{p}^{(k2)}$, and $\hat{T}_3$ observables are planewaves denoted in the ket notations by $|p^{(k1)}, p^{(k2)}, t_3\rangle$, where $(p^{(k1)}, p^{(k2)}, t_3)$ are the eigenvalues.
On the other hand, as in Section~\ref{subsec_hatK}, we can form a system of equations, using Eqs.~(\ref{def_Ti_op}) and (\ref{pk1_pk2_xyz}):
\begin{eqnarray}
\left[ \begin{array}{c}
\hat{p}^{(k1)} \\ \hat{p}^{(k2)} \\ \hat{T}_3
\end{array} \right] = \left[ \left[ T^{(k_1,k_2)}_3 \right] \right]
\left[ \begin{array}{c}
\hat{p}_1 \\ \hat{p}_2 \\ \hat{p}_3
\end{array} \right] ,
\qquad {\rm where} \qquad
\left[ \left[ T^{(k_1,k_2)}_3 \right] \right] \equiv \left[ \begin{array}{c}
\left( \bP^{(k1)} \right)^t \\ \left( \bP^{(k2)} \right)^t \\ (\bT_3^{(x,y,z)})^t
\end{array} \right]
\label{matrix_T3}
\end{eqnarray}
($\cdot^t$ denotes the transposed of a vector or a matrix), and which may be inverted, to obtain all the components of the momentum operator $\hat{\bp}$ in terms of $(\hat{p}^{(k1)}, \hat{p}^{(k2)}, \hat{T}_3)$ and $(k_1, k_2, u)$.
In terms of the variables $(k_1, k_2, u)$, the vectors that appear in Eq.~(\ref{matrix_T3}) are (see Eqs.~\ref{def_Ti_op} and \ref{xy_k1k2u_fxfy}-\ref{hat_pk1pk2_gen_expr_k1k2u}):
%
\begin{subequations} \label{defs_Pk1_Pk2_T3}
\begin{eqnarray}
    \bP^{(k1)} &\equiv& \left[ \begin{array}{c}
    \frac{\partial f_x(k_1, k_2, z)}{\partial k_1} + \frac{\partial f_x(k_1, k_2, z)}{\partial z} \frac{\partial z(k_1, k_2, u)}{\partial k_1} \\
    \frac{\partial f_y(k_1, k_2, z)}{\partial k_1} + \frac{\partial f_y(k_1, k_2, z)}{\partial z} \frac{\partial z(k_1, k_2, u)}{\partial k_1} \\
    \frac{\partial z(k_1,k_2,u)}{\partial k_1}
    \end{array}
    \right] , 
    \label{def_Pk1} \\
    \bP^{(k2)} &\equiv& \left[ \begin{array}{c}
    \frac{\partial f_x(k_1, k_2, z)}{\partial k_2} + \frac{\partial f_x(k_1, k_2, z)}{\partial z} \frac{\partial z(k_1, k_2, u)}{\partial k_2} \\
    \frac{\partial f_y(k_1, k_2, z)}{\partial k_2} + \frac{\partial f_y(k_1, k_2, z)}{\partial z} \frac{\partial z(k_1, k_2, u)}{\partial k_2} \\
    \frac{\partial z(k_1,k_2,u)}{\partial k_2}
    \end{array}
    \right] ,
    \label{def_Pk2} \\
    \bT_3^{(x,y,z)} &\equiv& \frac{1}{10mc} \left[\begin{array}{c}
    x(k_1,k_2,u) z(k_1,k_2,u) \\ y(k_1,k_2,u)z(k_1,k_2,u) \\ z^2(k_1,k_2,u) - 2r^2(k_1,k_2,u)
    \end{array}\right] .
    \label{def_T3}
\end{eqnarray}
\end{subequations}
%

\section{Discussion} \label{sec_discussion}

As we discussed in Section~\ref{subsec_Hilbert_sp}, the Hilbert space $\hH = L^2(\R^3)$ in the coordinates $(x,y,z)$ is mapped onto the Hilbert space $\tilde{\tilde{\hH}} = L^2(\M_1)$ in the coordinates $(k_1,k_2,u)$,  where $\hat{T}_3 = -i\hbar \partial/\partial u$.
In Ref.~\cite{JPA30.3515.1997.Anghel} it was shown that since no restrictions are imposed for the functions in $\tilde{\tilde{\hH}}$ on the boundaries of $\M_1$, then $\hat{T}_3$ is hypermaximal.
It can be shown, by using the same procedure as in Ref.~\cite{JPA30.3515.1997.Anghel}, but in the coordinates $(k_1,k_2,u)$, that also the operators $\hat{p}_{k1} = -i\hbar \partial/\partial k_1$ and $\hat{p}_{k2} = -i\hbar \partial/\partial k_2$ are hypermaximal.
Nevertheless, in physical systems (in nuclear physics ~\cite{SovPhysJETP.6.1184.1958.Zeldovich, Phys.Rev.C.65.1.2002.Haxton, AIPConfProc.477.14.1999.Flambaum}, solid state physics ~\cite{PhysRep.187.145.1990.Dubovik}, and metamaterials ~\cite{Nanotechnology.30.2019.Yang, Comm.Pres.2.10.2019.Savinov, NJP.9.324.2007.Marinov}), toroidal moments are not generated by currents corresponding  to the $\hat{T}_3$ eigenvectors, that is, currents along the continuous lines in Fig.~\ref{fig_coordinates}(a), but they are associated, in general, with currents distributions along the poloidal lines of tori, as depicted in Fig.~\ref{fig_tor}.
Then, the physical restrictions imposed on the concrete systems specify the boundary conditions.
For example, we may have Dirichlet boundary conditions for a particle moving in a restricted geometry, or periodic boundary conditions for a particle moving in a condensed matter system or a metamaterial.

In this section we shall consider a quantum particle in a region contained in a thin torus  of major radius $R_T$ and minor radius $r_T$, with $r_T \ll R_T$.
In this case, $z \ll \rho$ (cylindrical coordinates) or $z \ll \sqrt{x^2+y^2}$ (Cartesian coordinates) and the transformation between the $(x,y,z)$ coordinates and the \textit{natural coordinates} of $\hat{T}_3$ are considerably simplified.


\subsection{Toroidal \textit{natural coordinates} in a thin torus} \label{subsec_thin_tor}

Let us first analyze the system in the $(k,u,\rho)$ coordinates.
From Eq.~(\ref{def_nat_set}), we observe that when $z=0$, $k(\rho,z=0) = \rho$ and $u(\rho,z=0) = 0$.
Since $r_T \ll R_T$ and $z \ll \rho$, we can expand the expressions for the coordinates $k$ and $u$ in Taylor series in $z$, around $z=0$,
%
\begin{equation}
    k \approx \rho \left( 1 + \frac{z^2}{4 \rho^2} \right)
    \quad {\rm and} \quad
    u \approx - 10 mc \int_0^z \left( \frac{1}{2k^2} - \frac{t^4}{k^6} \right) dt
    = - 10 mc \left( \frac{z}{2k^2} - \frac{z^5}{5 k^6} \right) ,
\end{equation}
which, in the lowest order, gives
\begin{subequations}\label{ops_coord_tor_thin_cyl}
\begin{equation}
k = \rho
\qquad {\rm and} \qquad
u = - 5 mc \frac{z}{\rho^2} \approx - \frac{5 mc}{R_T^2} z.
\label{tor_thin_cyl}
\end{equation}
We see that inside the thin torus, the coordinates $(k,u,\phi)$ are just a re-scaling of the cylindrical coordinates $(\rho,z,\phi)$.
Similarly, the operators $\hat{p}^{(k)}$ and $\hat{T}_3$ have simple forms,
\begin{equation}
    \hat{p}^{(k)} = - i\hbar \frac{\partial}{\partial\rho}
    \qquad {\rm and} \qquad
    \hat{T}_3 = i\hbar \frac{R_T^2}{5 mc} \frac{\partial}{\partial z} . 
    \label{ops_tor_thin_cyl}
\end{equation}
\end{subequations}
Equations~(\ref{ops_coord_tor_thin_cyl}) are much more simple than Eqs.~(\ref{def_T3_op_cyl}), (\ref{def_nat_set}), and (\ref{def_pk}), so they may be straightforwardly inverted (omitting the procedure of  Section~\ref{subsec_hatK}).
By doing so, we can express the free particle Hamiltonian in the coordinates $(k,u,\phi)$, which take a very simple form in a thin torus:
\begin{equation}
    H_0 = - \frac{\hbar^2 \Delta}{2m} = - \frac{\hbar^2}{2m} \left[ \frac{1}{\rho} \frac{\partial}{\partial \rho} \left(\rho \frac{\partial}{\partial\rho}\right) + \frac{1}{\rho^2} \frac{\partial^2}{\partial\phi^2} + \frac{\partial^2}{\partial z^2}
    \right]
    = - \frac{\hbar^2}{2m} \left[ \frac{1}{k} \frac{\partial}{\partial k} \left(k \frac{\partial}{\partial k}\right) + \frac{1}{k^2} \frac{\partial^2}{\partial\phi^2} + \left(\frac{5 mc}{R_T^2}\right)^2 \frac{\partial^2}{\partial u^2}
    \right]
    \label{H0_k}
\end{equation}

In the variables $(k_1,k_2,u)$, from Eqs.~(\ref{defs_k1k2}) we obtain $k_1(x,y,z=0) = x$ and $k_2(x,y,z=0) = y$.
Therefore, the Taylor expansions of the coordinates $k_1$ and $k_2$ around $z=0$ give
\begin{eqnarray}
    && k_1 \approx x \left( 1 + \frac{1}{4} \frac{z^2}{x^2+y^2}\right) , \quad
    k_2 \approx y \left( 1 + \frac{1}{4} \frac{z^2}{x^2+y^2}\right) , \quad
    {\rm and} \nonumber \\
    && u \approx - 10 mc \int_0^z \left[ \frac{1}{2(k_1^2+k_2^2)} - \frac{t^4}{(k_1^2+k_2^2)^3} \right] dt
    = - 10 mc \left( \frac{z}{2 (k_1^2+k_2^2)} - \frac{z^5}{5 (k_1^2+k_2^2)^3} \right) .
    \nonumber
\end{eqnarray}
In the lowest order of expansion, the equations above lead to
\begin{eqnarray}
&& k_1 \approx x , \quad
k_2 \approx y , \quad
u \approx - 5 mc \frac{z}{R_T^2} , \nonumber \\
&& \hat{p}^{(k1)} = -i\hbar \frac{\partial}{\partial x} \equiv \hat{p}_1 ,
\quad
\hat{p}^{(k2)} = -i\hbar \frac{\partial}{\partial y} \equiv \hat{p}_2 ,
\quad {\rm and} \quad
\hat{T}_3 = i\hbar \frac{R_T^2}{5 mc} \frac{\partial}{\partial z} \equiv - \frac{R_T^2}{5 mc} \hat{p}_3 ,
\label{ops_tor_thin_k1k2}
\end{eqnarray}
whereas the free  particle Hamiltonian~(\ref{H0_k}) is
\begin{equation}
\hat{H_0} = - \frac{\hbar^2}{2m} \left[ \frac{\partial^2}{\partial k_1^2} + \frac{\partial^2}{\partial k_2^2} + \left(\frac{5 mc}{R_T^2}\right)^2 \frac{\partial^2}{\partial u^2}
\right].
\label{H0_k1k2}
\end{equation}
%


\section{Conclusions} \label{sec_conclusions}

We analyzed the quantum operator $\hat{T}_3$, which corresponds to the projection of the toroidal moment onto the $z$ axis.
We used two ``natural sets of coordinates'' for $\hat{T}_3$, introduced in~\cite{JPA30.3515.1997.Anghel} and denoted here by $(k,u,\phi)$ and $(k_1, k_2, u)$, where $\phi$ is the azimuthal angle, $k_1 = k \cos\phi$, and $k_2 = k \sin\phi$.
In both these sets of coordinates, $\hat{T}_3 = - i\hbar \partial / \partial u$, and, in the $(x,y,z)$ space, represents the derivatives along the continuous lines in Fig.~\ref{fig_coordinates}(a).
Therefore, all the operators which are functions of $k$, $\partial / \partial k$, $\partial / \partial u$, and $\partial / \partial \phi$ in the coordinates $(k,u,\phi)$, or of $k_1$, $k_2$, $\partial / \partial k_1$, $\partial / \partial k_2$, $\partial / \partial u$ in the coordinates $(k_1, k_2, u)$, commute with $\hat{T}_3$.
Of these operators we did not know $\partial / \partial k$, $\partial / \partial k_1$, and $\partial / \partial k_2$, so we find their expressions in terms of the $(x,y,z)$ coordinates and the partial derivatives $\partial/\partial x$, $\partial/\partial y$, and $\partial/\partial z$.
Then, we calculate the inverse transformations, to express $\rho$, $z$, $\partial/\partial\rho$, and $\partial/\partial z$ in the natural coordinates $k$, $u$, $\partial/\partial k$, and $\partial/\partial u$.
Similarly, we express $x$, $y$, $z$, $\partial/\partial x$, $\partial/\partial y$, and $\partial/\partial z$ in terms of $k_1$, $k_2$, $k_3$, $\partial/\partial k_1$, $\partial/\partial k_2$, and $\partial/\partial u$.
These transformations enables us to express all the familiar operators, like positions, momenta, and the Hamiltonian operator in terms of operators in the natural coordinates of $\hat{T}_3$.

Although the general expressions are complicated, in specific, physically relevant cases, they may be substantially simplified.
One such case is when the quantum particle exists in a thin torus.
This case is analyzed in Section~\ref{sec_discussion} and simple expressions are found for momenta and the free particle Hamiltonian in terms of $\hat{T}_3$, the natural coordinates of $\hat{T}_3$, plus $\hat{p}_k$, $\hat{p}_{k1}$, and $\hat{p}_{k2}$.
This simplifies considerably, for example, the calculations of the expectation values of toroidal moment operator and of other commuting observables, like $\hat{p}^{(k1)}$ and $\hat{p}^{(k2)}$, in eigenstates of the Hamiltonian.

\section{Acknowledgements} \label{sec_ack}

This work has been financially supported by UEFISCDI project PN-19060101/2019 and the ELI-RO contract 81-44 /2020.
Travel support from Romania-JINR collaboration projects positions 23, 26, Order 397/27.05.2019 is gratefully acknowledged.

\appendix

\section{The calculation of $\hat p^{(k)}$} \label{app_pk}

In Eqs.~(\ref{def_nat_set}) we denote
\begin{equation}
f_{u}\left(z, k\right) \equiv -10mc\int_{0}^{z} \frac{d t}{\sqrt{t^{4}+4 k^{4}}}
\end{equation}
and write
\begin{subequations} \label{diffs_uk}
\begin{eqnarray}
d u &=& \frac{\partial f_u}{\partial z} d z+\frac{\partial f_u}{\partial k} d k , \label{diff_u1} \\
dk &=& \frac{\partial k}{\partial \rho} d \rho+\frac{\partial k}{\partial z} d z , \label{diff_k}
\end{eqnarray}
\end{subequations}
where
\begin{subequations} \label{part_derivs_uk}
\begin{eqnarray}
\frac{\partial f_{u}}{\partial z} &=& \frac{-10mc}{\sqrt{z^{4}+4 k^{4}}}
\equiv \frac{-10mc}{\sqrt{z^{4}+4 \left[ \rho^2 \left( z^2 + \rho^2 \right) \right]}}
\label{part_fu_z} \\
\frac{\partial f_u}{\partial k} &=& 5 m c \int_{0}^{z} \frac{16 k^{3}}{\left(t^{4}+4 k^{4}\right)^{3 / 2}} d t
\equiv 5 m c \int_{0}^{z} \frac{16 \left[ \rho^2 \left( z^2 + \rho^2 \right) \right]^{3/4}}{\left\{t^{4}+4 \left[ \rho^2 \left( z^2 + \rho^2 \right) \right]\right\}^{3 / 2}} d t
\label{part_fu_k} \\
\frac{\partial k}{\partial \rho} &=& \frac {2\,{\rho}^{2}+{z}^{2}}{2 \sqrt{\rho} \left( {\rho}^{2}+{z}^{2} \right) ^{3/4}}
\label{part_k_rho} \\
\frac{\partial k}{\partial z} &=& \frac {\sqrt {\rho}z}{2 \left( {\rho}^{2}+{z}^{2} \right) ^{3/4}}
\label{part_k_z}
\end{eqnarray}
\end{subequations}
From the condition $du = 0$ and Eq.~(\ref{diff_u1}) we obtain the derivative of $z$ with respect to $k$ at constant $u$:
\begin{equation}
\left.\frac{d z}{d k}\right|_{u} = - \frac{\partial f_u}{\partial k} \left(\frac{\partial f_u}{\partial z}\right)^{-1}
\label{dk_dz_uc}
\end{equation}

To obtain the derivative of $\rho$ with respect to $z$ at constant $u$, we plug Eq.~(\ref{dk_dz_uc}) into Eq.~(\ref{diff_k}) and, using also Eqs.~(\ref{part_derivs_uk}), we obtain
\begin{eqnarray}
    \left. \frac{d\rho}{dz} \right|_{u} &=& \frac{(dk/dz)_u - \partial k/\partial z}{\partial k/ \partial \rho}
    = - \frac{\frac{\partial f_u}{\partial z} \left(\frac{\partial f_u}{\partial k}\right)^{-1} + \frac {\sqrt {\rho}z}{2 \left( {\rho}^{2}+{z}^{2} \right) ^{3/4}}}{\frac {2\,{\rho}^{2}+{z}^{2}}{2 \sqrt{\rho} \left( {\rho}^{2}+{z}^{2} \right) ^{3/4}}}
    \label{drho_dz_uct}
\end{eqnarray}
Combining~(\ref{drho_dz_uct}) with~(\ref{dk_dz_uc}), we write
\begin{subequations} \label{derivs_for_pk}
\begin{equation}
    \left. \frac{d\rho}{dk} \right|_{u} = \left. \frac{d\rho}{dz} \right|_{u} \left. \frac{dz}{d k} \right|_{u}
    = - \left. \frac{d\rho}{dz} \right|_{u}
    \frac{\partial f_u}{\partial k} \left(\frac{\partial f_u}{\partial z}\right)^{-1}
    \label{dz_du_dk}
\end{equation}
Then, the partial derivative in the $(k,u,\phi)$ space of an arbitrary function $\tilde{F}(k,u,\phi) \equiv F(\rho,z,\phi)$ may be written in cylindrical coordinates as
\begin{equation}
    \frac{\partial \tilde{F}(k,u,\phi)}{\partial k}
    = \frac{\partial F}{\partial \rho} \left. \frac{d\rho}{dk} \right|_{u} + \frac{\partial F}{\partial z} \left. \frac{dz}{dk} \right|_{u}
\end{equation}
\end{subequations}
Combining Eqs.~(\ref{part_derivs_uk}) and (\ref{derivs_for_pk}) we obtain the expression for $\hat{p}^{(k)}$:
\begin{eqnarray}
    \hat{p}^{(k)} &\equiv& - i\hbar \frac{\partial}{\partial k}
    = - i\hbar \left( \left. \frac{d\rho}{dk} \right|_{u} \frac{\partial}{\partial \rho} + \left. \frac{dz}{dk} \right|_{u} \frac{\partial}{\partial z} \right)
    \label{pk_fin_app}
\end{eqnarray}

On the other hand, if we invert Eqs.~(\ref{def_nat_set}), by first formally writing from the second equation $z$ as a function of $k$ and $u$ (say, $z(k,u)$, since we do not have an analytical expression), then, from  the first equation,
\begin{equation}
    \rho (k,u) = \frac{z(k,u)}{\sqrt{2}} \sqrt{\sqrt{1 + 4 \frac{k^4}{z^4(k,u)}} - 1} .
    \label{rho_ku}
\end{equation}
In these notations,
\begin{equation}
    \left. \frac{dz}{dk} \right|_{u} \equiv \frac{\partial z(k,u)}{\partial k},
    \quad {\rm and} \quad
    \left. \frac{d\rho}{dk} \right|_{u} \equiv \frac{\partial \rho(k,u)}{\partial k}
    = - \frac {\sqrt{\sqrt {1 + 4{k}^{4}/{z}^{4}} - 1}}{ \sqrt{2} \sqrt{1 + 4{k}^{4}/{z}^{4}}}
    \frac{\partial z}{\partial k}
    + \frac {2\sqrt {2}{k}^{3} / z^3}{ \sqrt {1 + \sqrt{4{k}^{4}/{z}^{4} } - 1 } \sqrt{1 + 4{k}^{4}/{z}^{4}}} ,
    \label{derivs_zrho_k}
\end{equation}
which may be plugged into~(\ref{pk_fin_app}), as it will be used in  Section~\ref{sec_discussion}.

\section{The calculation of $\hat p^{(k1)}$ and $\hat p^{(k2)}$} \label{app_pk1pk2}

From Eqs.~(\ref{def_nat_set_k1k2}) and (\ref{defs_k1k2}) we write
\begin{subequations} \label{diffs_uk1k2}
    \begin{eqnarray}
    du &=& \frac{\partial f_{u2}}{\partial z} dz + \frac{\partial f_{u2}}{\partial k_1} dk_1 + \frac{\partial f_{u2}}{\partial k_2} dk_2 , \label{diff_u} \\
    dk_1 &=& \frac{\partial k_1}{\partial x} dx + \frac{\partial k_1}{\partial y} dy + \frac{\partial k_1}{\partial z} dz , \label{diff_k1} \\
    dk_2 &=& \frac{\partial k_2}{\partial x} dx + \frac{\partial k_2}{\partial y} dy + \frac{\partial k_2}{\partial z} dz , \label{diff_k2}
    \end{eqnarray}
\end{subequations}
where
\begin{subequations} \label{part_derivs}
    \begin{eqnarray}
    \frac{\partial f_{u2}}{\partial z}   &=& \frac{- 10 mc}{\sqrt{z^4 + 4 (k_1^2 + k_2^2)^2}}
    \label{part_fu2_z} \\
    \frac{\partial f_{u2}}{\partial k_1} &=& 80 mc \left( {k_1}^{2}+{k_2}^{2} \right) k_1 \int_0^z \frac { dt }{ \left[ {t}^{4}+4 \left( {k_1}^{2}+{k_2}^{2} \right) ^{2} \right] ^{3/2}} ,
    \label{part_fu2_k1} \\
    \frac{\partial f_{u2}}{\partial k_2} &=& 80 mc \left( {k_1}^{2}+{k_2}^{2} \right) k_2 \int_0^z \frac { dt }{ \left[ {t}^{4}+4 \left( {k_1}^{2}+{k_2}^{2} \right) ^{2} \right] ^{3/2}} ,
    \label{part_fu2_k2} \\
    && {k_1}^{2}+{k_2}^{2} = \rho r = \sqrt{(x^2 + y^2)(x^2 + y^2 + z^2)} \nonumber \\
    \frac{\partial k_1}{\partial x} &=& \frac{4 {x}^{2} {y}^{2} + {x}^{2} {z}^{2} + 2 {y}^{2}{z}^{2} + 2 {x}^{4} + 2 {y}^{4}}{ 2 \left( {x}^{2}+{y}^{2} \right) ^{5/4} \left( {x}^{2}+{y}^{2}+{z}^{2} \right) ^{3/4}} ,
    \label{dk1_dx} \\
    \frac{\partial k_1}{\partial y} &=& - \frac {xy{z}^{2}}{ 2 \left( {x}^{2}+{y}^{2} \right) ^{5/4} \left( {x}^{2}+{y}^{2}+{z}^{2} \right) ^{3/4}} ,
    \label{dk1_dy} \\
    \frac{\partial k_1}{\partial z} &=& \frac {xz}{ 2 \left( {x}^{2}+{y}^{2} \right)^{1/4} \left( {x}^{2}+{y}^{2}+{z}^{2} \right)^{3/4}} ,
    \label{dk1_dz} \\
    \frac{\partial k_2}{\partial x} &=& - \frac{xy{z}^{2}}{ 2 \left( {x}^{2}+{y}^{2} \right) ^{5/4} \left( {x}^{2}+{y}^{2}+{z}^{2} \right) ^{3/4}} ,
    \label{dk2_dx} \\
    \frac{\partial k_2}{\partial y} &=& \frac {4 {x}^{2}{y}^{2}+2 {x}^{2}{z}^{2}+ {y}^{2}{z}^{2} + 2 {x}^{4}+ 2{y}^{4} }{ 2 \left( {x}^{2}+{y}^{2} \right) ^{5/4} \left( {x}^{2} + {y}^{2}+{z}^{2} \right) ^{3/4}} ,
    \label{dk2_dy} \\
    \frac{\partial k_2}{\partial z} &=& \frac{yz}{ 2 ({x}^{2}+{y}^{2})^{1/4} \left( {x}^{2}+{y}^{2}+{z}^{2} \right) ^{3/4}} .
    \label{dk2_dz}
    \end{eqnarray}
\end{subequations}

To express $\partial/\partial k_1$ in the $(x,y,z)$ coordinates, we impose $du  = dk_2 = 0$ in Eq.~(\ref{diff_u}) and obtain
\begin{equation}
%
dz = - \frac{\partial f_{u2}}{\partial k_1} \left(\frac{\partial f_{u2}}{\partial z}\right)^{-1} dk_1 . \label{dk11}
\end{equation}
Plugging~(\ref{dk11}) into~(\ref{diff_k1}) we obtain
\begin{equation}
- \left[ \frac{\partial f_{u2}}{\partial z} \left(\frac{\partial f_{u2}}{\partial k_1}\right)^{-1} + \frac{\partial k_1}{\partial z} \right] dz = \frac{\partial k_1}{\partial x} dx + \frac{\partial k_1}{\partial y} dy , \label{dx_dy_dz}
\end{equation}
whereas from~(\ref{diff_k2}) and $dk_2 = 0$ we obtain
\begin{equation}
dy = - \left(\frac{\partial k_2}{\partial y}\right)^{-1} \left( \frac{\partial k_2}{\partial x} dx + \frac{\partial k_2}{\partial z} dz \right) . \label{dy_dxdz}
\end{equation}
Plugging~(\ref{dy_dxdz}) into (\ref{dx_dy_dz}) we obtain an equation in $dx$ and $dz$,
\begin{eqnarray}
&& dx =
\frac{\frac{\partial k_1}{\partial y} \frac{\partial k_2}{\partial z} \left(\frac{\partial k_2}{\partial y}\right)^{-1} - \frac{\partial f_{u2}}{\partial z} \left(\frac{\partial f_{u2}}{\partial k_1}\right)^{-1} - \frac{\partial k_1}{\partial z}}{\frac{\partial k_1}{\partial x} - \frac{\partial k_1}{\partial y} \frac{\partial k_2}{\partial x} \left(\frac{\partial k_2}{\partial y}\right)^{-1}} dz
\label{dx_dz}
\end{eqnarray}
Plugging~(\ref{dx_dz}) into (\ref{dy_dxdz}) we obtain
\begin{eqnarray}
    dy = - \left(\frac{\partial k_2}{\partial y}\right)^{-1} \left[ \frac{\partial k_2}{\partial x} \frac{\frac{\partial k_1}{\partial y} \frac{\partial k_2}{\partial z} \left(\frac{\partial k_2}{\partial y}\right)^{-1} - \frac{\partial f_{u2}}{\partial z} \left(\frac{\partial f_{u2}}{\partial k_1}\right)^{-1} - \frac{\partial k_1}{\partial z}}{\frac{\partial k_1}{\partial x} - \frac{\partial k_1}{\partial y} \frac{\partial k_2}{\partial x} \left(\frac{\partial k_2}{\partial y}\right)^{-1}} + \frac{\partial k_2}{\partial z} \right] dz
    \label{dy_dz}
\end{eqnarray}
Equations~(\ref{dk11}), (\ref{dx_dz}), and (\ref{dy_dz}) give us the derivatives at constant $u$ and  $k_2$:
\begin{eqnarray}
    && \left. \frac{dz}{dk_1}\right|_{u, k_2} = - \frac{\partial f_{u2}}{\partial k_1} \left(\frac{\partial f_{u2}}{\partial z}\right)^{-1} ,
    \qquad
    \left. \frac{dx}{dz}\right|_{u, k_2} = \frac{\frac{\partial k_1}{\partial y} \frac{\partial k_2}{\partial z} \left(\frac{\partial k_2}{\partial y}\right)^{-1} - \frac{\partial f_{u2}}{\partial z} \left(\frac{\partial f_{u2}}{\partial k_1}\right)^{-1} - \frac{\partial k_1}{\partial z}}{\frac{\partial k_1}{\partial x} - \frac{\partial k_1}{\partial y} \frac{\partial k_2}{\partial x} \left(\frac{\partial k_2}{\partial y}\right)^{-1}} ,
    \quad {\rm and} \nonumber \\
    && \left. \frac{dy}{dz}\right|_{u, k_2} = - \left(\frac{\partial k_2}{\partial y}\right)^{-1} \left[ \frac{\partial k_2}{\partial x} \frac{\frac{\partial k_1}{\partial y} \frac{\partial k_2}{\partial z} \left(\frac{\partial k_2}{\partial y}\right)^{-1} - \frac{\partial f_{u2}}{\partial z} \left(\frac{\partial f_{u2}}{\partial k_1}\right)^{-1} - \frac{\partial k_1}{\partial z}}{\frac{\partial k_1}{\partial x} - \frac{\partial k_1}{\partial y} \frac{\partial k_2}{\partial x} \left(\frac{\partial k_2}{\partial y}\right)^{-1}} + \frac{\partial k_2}{\partial z} \right]
    \label{derivs_u_k2_ct}
\end{eqnarray}
Finally, from~(\ref{derivs_u_k2_ct}) we obtain the expression for $\hat{p}^{(k1)}$
\begin{eqnarray}
    \hat{p}^{(k1)} &\equiv& - i\hbar \frac{\partial}{\partial k_1}
    = - i\hbar \left. \frac{dz}{dk_1}\right|_{u,k_2} \left\{ \left. \frac{dx}{dz}\right|_{u, k_2} \frac{\partial}{\partial x}
    + \left. \frac{dy}{dz}\right|_{u, k_2} \frac{\partial}{\partial y}
    + \frac{\partial}{\partial z} \right\} .
    \label{hat_pk1_gen_expr}
\end{eqnarray}

Similarly, we can obtain the expression for $\hat{p}^{(k2)}$.
From Eq.~(\ref{diff_u}), with the conditions $du = dk_1 = 0$, we get
\begin{equation}
dz = - \frac{\partial f_{u2}}{\partial k_2} \left(\frac{\partial f_{u2}}{\partial z}\right)^{-1} dk_2 . \label{dk21}
\end{equation}
Plugging~(\ref{dk21}) into~(\ref{diff_k2}) we obtain
\begin{equation}
- \left[ \frac{\partial f_{u2}}{\partial z} \left(\frac{\partial f_{u2}}{\partial k_2}\right)^{-1} + \frac{\partial k_2}{\partial z} \right] dz = \frac{\partial k_2}{\partial x} dx + \frac{\partial k_2}{\partial y} dy , \label{dx_dy_dz2}
\end{equation}
whereas from~(\ref{diff_k1}) and $dk_1 = 0$ we have
\begin{equation}
dy = - \left(\frac{\partial k_1}{\partial y}\right)^{-1} \left( \frac{\partial k_1}{\partial x} dx + \frac{\partial k_1}{\partial z} dz \right) . \label{dy_dxdz2}
\end{equation}
Plugging~(\ref{dy_dxdz2}) into (\ref{dx_dy_dz2}) we obtain
\begin{eqnarray}
&& dx =
\frac{\frac{\partial k_2}{\partial y} \frac{\partial k_1}{\partial z} \left(\frac{\partial k_1}{\partial y}\right)^{-1} - \frac{\partial f_{u2}}{\partial z} \left(\frac{\partial f_{u2}}{\partial k_2}\right)^{-1} - \frac{\partial k_2}{\partial z}}
{\frac{\partial k_2}{\partial x} - \frac{\partial k_2}{\partial y} \frac{\partial k_1}{\partial x} \left(\frac{\partial k_1}{\partial y}\right)^{-1}} dz
\label{dx_dz2}
\end{eqnarray}
Plugging~(\ref{dx_dz2}) into (\ref{dy_dxdz2}) we obtain
\begin{eqnarray}
dy = - \left(\frac{\partial k_1}{\partial y}\right)^{-1} \left[ \frac{\partial k_1}{\partial x}
\frac{\frac{\partial k_2}{\partial y} \frac{\partial k_1}{\partial z} \left(\frac{\partial k_1}{\partial y}\right)^{-1} - \frac{\partial f_{u2}}{\partial z} \left(\frac{\partial f_{u2}}{\partial k_2}\right)^{-1} - \frac{\partial k_2}{\partial z}}
{\frac{\partial k_2}{\partial x} - \frac{\partial k_2}{\partial y} \frac{\partial k_1}{\partial x} \left(\frac{\partial k_1}{\partial y}\right)^{-1}}
  + \frac{\partial k_1}{\partial z} \right] dz
\label{dy_dz2}
\end{eqnarray}
From Eqs.~(\ref{dk21}), (\ref{dx_dz2}), and (\ref{dy_dz2}) give us the derivatives at constant $u$ and $k_1$:
\begin{eqnarray}
    && \left. \frac{dz}{dk_2} \right|_{u,k_1} = - \frac{\partial f_{u2}}{\partial k_2} \left(\frac{\partial f_{u2}}{\partial z}\right)^{-1} ,
    \quad
    \left. \frac{dx}{dz} \right|_{u,k_1} =
    \frac{\frac{\partial k_2}{\partial y} \frac{\partial k_1}{\partial z} \left(\frac{\partial k_1}{\partial y}\right)^{-1} - \frac{\partial f_{u2}}{\partial z} \left(\frac{\partial f_{u2}}{\partial k_2}\right)^{-1} - \frac{\partial k_2}{\partial z}}
    {\frac{\partial k_2}{\partial x} - \frac{\partial k_2}{\partial y} \frac{\partial k_1}{\partial x} \left(\frac{\partial k_1}{\partial y}\right)^{-1}}
    \nonumber \\
    && \left. \frac{dy}{dz} \right|_{u,k_1} = - \left(\frac{\partial k_1}{\partial y}\right)^{-1} \left[ \frac{\partial k_1}{\partial x}
    \frac{\frac{\partial k_2}{\partial y} \frac{\partial k_1}{\partial z} \left(\frac{\partial k_1}{\partial y}\right)^{-1} - \frac{\partial f_{u2}}{\partial z} \left(\frac{\partial f_{u2}}{\partial k_2}\right)^{-1} - \frac{\partial k_2}{\partial z}}
    {\frac{\partial k_2}{\partial x} - \frac{\partial k_2}{\partial y} \frac{\partial k_1}{\partial x} \left(\frac{\partial k_1}{\partial y}\right)^{-1}}
    + \frac{\partial k_1}{\partial z} \right]
    \label{derivs_u_k1_ct}
\end{eqnarray}

we obtain the expression for the $\hat{p}^{(k2)}$
\begin{eqnarray}
\hat{p}^{(k2)} &\equiv& - i\hbar \frac{\partial}{\partial k_2}
= - i\hbar \left. \frac{dz}{dk_2}\right|_{u,k_1} \left\{ \left. \frac{dx}{dz}\right|_{u, k_1} \frac{\partial}{\partial x}
+ \left. \frac{dy}{dz}\right|_{u, k_1} \frac{\partial}{\partial y}
+ \frac{\partial}{\partial z} \right\} .
\label{hat_pk2_gen_expr} \\
\end{eqnarray}

As it was done in Appendix~\ref{app_pk}, it will be useful to express the derivatives with respect to $k_1$ and $k_2$ in the variables $(k_1, k_2, u)$.
For this, we formally invert Eqs.~(\ref{def_nat_set_k1k2}), to write $z$ as $z(k_1, k_2, u)$, a function of $k_1$, $k_2$, and $u$.
Then, from Eqs.~(\ref{defs_k1k2}), we obtain
\begin{subequations} \label{xy_k1k2u_fxfy}
\begin{equation}
    x(k_1, k_2, u) = f_x[k_1, k_2, z(k_1, k_2, u)]
    \qquad {\rm and} \qquad
    y(k_1, k_2, u) = f_y[k_1, k_2, z(k_1, k_2, u)] ,
    \label{xy_k1k2u}
\end{equation}
where
\begin{equation}
f_x(k_1, k_2, z) = \frac{k_1 z}{\sqrt{k_1^2 + k_2^2}} \sqrt{\sqrt{1 + 4 \frac{(k_1^2+k_2^2)^2}{z^4}} - 1}
\quad {\rm and} \quad
f_y(k_1, k_2, z) = \frac{k_2 z}{\sqrt{k_1^2 + k_2^2}} \sqrt{\sqrt{1 + 4 \frac{(k_1^2+k_2^2)^2}{z^4}} - 1}.
\label{def_fxfy}
\end{equation}
\end{subequations}
From the definition of $z(k_1, k_2, u)$, we write formally,
\begin{equation}
    \left. \frac{dz}{dk_1}\right|_{u,k_2} = \frac{\partial z(k_1, k_2, u)}{\partial k_1},
    \qquad
    \left. \frac{dz}{dk_2}\right|_{u,k_1} = \frac{\partial z(k_1, k_2, u)}{\partial k_2},
    \label{derivs_z_k1_k2}
\end{equation}
whereas the other derivatives are
\begin{subequations} \label{dfx_dfy_dk1k2z}
\begin{eqnarray}
    \frac{\partial f_x(k_1, k_2, z)}{\partial k_1} &=& \frac { k_{2}^{2}z }{ \left( k_{1}^{2}+k_{2}^{2} \right) ^{3/2}}
    \frac{ - \sqrt{ 1 + \frac{4 (k_{1}^2 + k_{2}^{2})^2 }{{z}^{4}}} + 1 + 4 \frac{( k_{1}^2 + k_{2}^2)^3}{ k_{2}^{2}{z}^{4}}}{
    \sqrt {1 + \frac{4 (k_{1}^{4} + k_{2}^2)^2 }{{z}^{4}}}
    \sqrt{\sqrt {1 + \frac{4 (k_{1}^{4} + k_{2}^2)^2 }{{z}^{4}}}-1}
    }
    \label{dfx_dk1} \\
    \frac{\partial f_x(k_1, k_2, z)}{\partial k_2} &=& \frac{k_{1} k_{2} z}{ \left( k_{1}^{2} + k_{2}^{2} \right)^{3/2}}
    \frac{\sqrt{ \sqrt{1 + {\frac {4 (k_{1}^2 + k_{2}^2)^2 }{{z}^{4}}}}-1}}{\sqrt{1 + {\frac {4 (k_{1}^2 + k_{2}^2)^2 }{{z}^{4}}}}}
    \label{dfx_dk2} \\
    \frac{\partial f_x(k_1, k_2, z)}{\partial z}
    %
    &=& - \frac{k_{1}}{\sqrt{{k_{1}}^{2}+{k_{2}}^{2}}}
    \left[ 1- \frac{1}{\sqrt{ \sqrt{1 + \frac{4 (k_{1}^2 + k_{2}^2)^2}{{z}^{4}}}}}\right]
    \label{dfx_dz} \\
    \frac{\partial f_y(k_1, k_2, z)}{\partial k_1} &=& \frac{k_{1} k_{2} z}{ \left( k_{1}^{2} + k_{2}^{2} \right)^{3/2}}
    \frac{\sqrt{ \sqrt{1 + {\frac {4 (k_{1}^2 + k_{2}^2)^2 }{{z}^{4}}}}-1}}{\sqrt{1 + {\frac {4 (k_{1}^2 + k_{2}^2)^2 }{{z}^{4}}}}}
    \label{dfy_dk1} \\
    \frac{\partial f_y(k_1, k_2, z)}{\partial k_2} &=& \frac { k_{1}^{2}z }{ \left( k_{1}^{2}+k_{2}^{2} \right) ^{3/2}}
    \frac{ - \sqrt{ 1 + \frac{4 (k_{1}^2 + k_{2}^{2})^2 }{{z}^{4}}} + 1 + 4 \frac{( k_{1}^2 + k_{2}^2)^3}{ k_{1}^{2}{z}^{4}}}{
    \sqrt {1 + \frac{4 (k_{1}^{4} + k_{2}^2)^2 }{{z}^{4}}}
    \sqrt{\sqrt {1 + \frac{4 (k_{1}^{4} + k_{2}^2)^2 }{{z}^{4}}}-1}
    }
    \label{dfy_dk2} \\
    \frac{\partial f_y(k_1, k_2, z)}{\partial z} &=& \frac{k_{2}}{\sqrt{{k_{1}}^{2}+{k_{2}}^{2}}}
    \left[ 1- \frac{1}{\sqrt{ \sqrt{1 + \frac{4 (k_{1}^2 + k_{2}^2)^2}{{z}^{4}}}}}\right]
    \label{dfy_dz}
\end{eqnarray}
\end{subequations}
Plugging Eqs.~(\ref{xy_k1k2u_fxfy})-(\ref{dfx_dfy_dk1k2z}) into the expressions for $\hat{p}^{(k1)}$ and $\hat{p}^{(k2)}$,  we obtain
\begin{subequations} \label{hat_pk1pk2_gen_expr_k1k2u}
\begin{eqnarray}
\hat{p}^{(k1)} &\equiv& - i\hbar \frac{\partial}{\partial k_1}
= - i\hbar \left\{ \frac{\partial x(k_1,k_2,u)}{\partial k_1} \frac{\partial}{\partial x}
+ \frac{\partial y(k_1,k_2,u)}{\partial k_1} \frac{\partial}{\partial y}
+ \frac{\partial z(k_1,k_2,u)}{\partial k_1} \frac{\partial}{\partial z} \right\} \nonumber \\
&=& - i\hbar \left\{ \left[ \frac{\partial f_x(k_1, k_2, z)}{\partial k_1} + \frac{\partial f_x(k_1, k_2, z)}{\partial z} \frac{\partial z(k_1, k_2, u)}{\partial k_1} \right] \frac{\partial}{\partial x} \right. \nonumber \\
&& \left. + \left[ \frac{\partial f_y(k_1, k_2, z)}{\partial k_1} + \frac{\partial f_y(k_1, k_2, z)}{\partial z} \frac{\partial z(k_1, k_2, u)}{\partial k_1} \right] \frac{\partial}{\partial y}
+ \frac{\partial z(k_1,k_2,u)}{\partial k_1} \frac{\partial}{\partial z} \right\}
\label{hat_pk1_gen_expr_k1k2u}
\end{eqnarray}
and
\begin{eqnarray}
\hat{p}^{(k2)} &\equiv& - i\hbar \frac{\partial}{\partial k_2}
= - i\hbar \left\{ \frac{\partial x(k_1,k_2,u)}{\partial k_2} \frac{\partial}{\partial x}
+ \frac{\partial y(k_1,k_2,u)}{\partial k_2} \frac{\partial}{\partial y}
+ \frac{\partial z(k_1,k_2,u)}{\partial k_2} \frac{\partial}{\partial z} \right\} \nonumber \\
&=& - i\hbar \left\{ \left[ \frac{\partial f_x(k_1, k_2, z)}{\partial k_2} + \frac{\partial f_x(k_1, k_2, z)}{\partial z} \frac{\partial z(k_1, k_2, u)}{\partial k_2} \right] \frac{\partial}{\partial x} \right. \nonumber \\
&& \left. + \left[ \frac{\partial f_y(k_1, k_2, z)}{\partial k_2} + \frac{\partial f_y(k_1, k_2, z)}{\partial z} \frac{\partial z(k_1, k_2, u)}{\partial k_2} \right] \frac{\partial}{\partial y}
+ \frac{\partial z(k_1,k_2,u)}{\partial k_2} \frac{\partial}{\partial z} \right\} .
\label{hat_pk2_gen_expr_k1k2u}
\end{eqnarray}
\end{subequations}

\end{document}

%% file: color.tex
\usepackage[usenames]{color}

\definecolor{brown}{rgb}{0.3,0.2,0}

%% file: talk.tex
\usepackage{graphicx,bm,amssymb,amsmath}
\usepackage{ulem} 

\newcommand{\bj}{{\bf j}}

\newcommand{\bp}{{\bf p}}

\newcommand{\br}{{\bf r}}

\newcommand{\bt}{{\bf t}}

\newcommand{\bz}{{\bf z}}

\newcommand{\bA}{{\bf A}}

\newcommand{\bD}{{\bf D}}

\newcommand{\bH}{{\bf H}}

\newcommand{\bJ}{{\bf J}}

\newcommand{\bP}{{\bf P}}

\newcommand{\bS}{{\bf S}}
\newcommand{\bT}{{\bf T}}

\newcommand{\cH}{{\cal H}}

\newcommand{\cT}{{\cal T}}

\newcommand{\D}{\mathbb{D}}
\newcommand{\hH}{\mathbb{H}}

\newcommand{\M}{\mathbb{M}}

\newcommand{\R}{\mathbb{R}}

\def\tensor#1{\protect\@ontopof{#1}{\leftrightarrow}{1.15}\mathord{\box2}}




%% file: Tor_self_adjoint_v4_1.bbl
\begin{thebibliography}{10}
    
    \bibitem{SovPhysJETP.6.1184.1958.Zeldovich}
    Ya.~B. Zeldovich.
    \newblock Electromagnetic interaction with parity violation.
    \newblock {\em Sov. Phys.-JETP}, 6:1184, 1958.
    
    \bibitem{Sov.24.1965.Dubovik}
    V~M Dubovik and A~A Cheshkov.
    \newblock Form-factors and multipoles in electromagnetic interactions.
    \newblock {\em Sov. Phys. JETP}, 24, 1965.
    
    \bibitem{SovJ.5.318.1974.Dubovik}
    V~M Dubovik and A~A Cheshkov.
    \newblock Multipole expansion in classical and quantum field theory and
    radiation.
    \newblock {\em Sov. J. Particles Nucl.}, 5.
    
    \bibitem{nanz:book.2016toroidal}
    S.~Nanz.
    \newblock {\em Toroidal Multipole Moments in Classical Electrodynamics: An
        Analysis of their Emergence and Physical Significance}.
    \newblock BestMasters. Springer Fachmedien Wiesbaden, 2016.
    
    \bibitem{Science.275.1759.1997.Wood}
    C.~S. Wood, S.~C. Bennett, D.~Cho, B.~P. Masterson, J.~L. Roberts, C.~E.
    Tanner, and C.~E. Wieman.
    \newblock Measurement of parity nonconservation and an anapole moment in
    cesium.
    \newblock {\em Science}, 275(5307):1759, 1997.
    
    \bibitem{NatureMat.15.263.2016.Papasimakis}
    Nikitas Papasimakis, V.A. Fedotov, Vassili Savinov, T.A. Raybould, and N.I.
    Zheludev.
    \newblock Electromagnetic toroidal excitations in matter and free space.
    \newblock {\em Nature Mat.}, 15:263, 2016.
    
    \bibitem{AIPConfProc.477.14.1999.Flambaum}
    V.~V. Flambaum.
    \newblock Nuclear anapole moment and tests of the standard model.
    \newblock {\em AIP Conference Proceedings}, 477(1):14, 1999.
    
    \bibitem{PhysRev.70.965.1946.Kittel}
    Charles Kittel.
    \newblock Theory of the structure of ferromagnetic domains in films and small
    particles.
    \newblock {\em Phys. Rev.}, 70:965--971, 1946.
    
    \bibitem{Physics.2.20.2009.Khomskii}
    Daniel Khomskii.
    \newblock Classifying multiferroics: Mechanisms and effects.
    \newblock {\em Physics}, 2:20, 2009.
    
    \bibitem{PU.55.557.2012.Pyatakov}
    Aleksandr Pyatakov and Anatoly Zvezdin.
    \newblock Magnetoelectric and multiferroic media.
    \newblock {\em Physics-Uspekhi}, 55:557, 2012.
    
    \bibitem{JExpTPL.52.161.1990.Tolstoi}
    N.~Tolstoi and A.~Spartakov.
    \newblock Aromagnetism: A new type of magnetism.
    \newblock {\em J. Exp. Theor. Phys. Lett.}, 52:161, 01 1990.
    
    \bibitem{NewJPhys.9.95.2007.Fedotov}
    V.A. Fedotov, K.~Marinov, Allan Boardman, and N.I. Zheludev.
    \newblock On the aromagnetism and anapole moment of anthracene nanocrystals.
    \newblock {\em New J. Phys.}, 9:95, 04 2007.
    
    \bibitem{PhysRevB.84.094421.2011.Toledano}
    Pierre Tol\'edano, Dmitry Khalyavin, and Laurent Chapon.
    \newblock Spontaneous toroidal moment and field-induced magnetotoroidic effects
    in { $Ba_ {2} CoGe_ {2} O_ {7}$}.
    \newblock {\em Phys. Rev. B}, 84:094421, 2011.
    
    \bibitem{PhysRevB.101.2020.Shimada}
    Takahiro Shimada, Yuuki Ichiki, Gen Fujimoto, Le~Lich, Tao Xu, Jie Wang, and
    Hiroyuki Hirakata.
    \newblock Ferrotoroidic polarons in antiferrodistortive {$ SrTi O _{3}$}.
    \newblock {\em Phys. Rev. B}, 101, 2020.
    
    \bibitem{JexpThPhys.87.146.1998.Popov}
    Yu~Popov, A.~Kadomtseva, G.~Vorob’ev, VA~Timofeeva, D.~Ustinin, Anatoly
    Zvezdin, and M.~Tegeranchi.
    \newblock Magnetoelectric effect and toroidal ordering in
    {$Ga_{2-x}Fe_{x}O_{3}$}.
    \newblock {\em Journal of Experimental and Theoretical Physics}, 87:146, 07
    1998.
    
    \bibitem{NatNano.14.141.2019.Lehmann}
    Jannis Lehmann, Claire Donnelly, Peter Derlet, Laura Heyderman, and Manfred
    Fiebig.
    \newblock Poling of an artificial magneto-toroidal crystal.
    \newblock {\em Nature Nanotechnology}, 14:141, 2019.
    
    \bibitem{Nanotechnology.30.2019.Yang}
    Yuanqing Yang and Sergey Bozhevolnyi.
    \newblock Nonradiating anapole states in nanophotonics: from fundamentals to
    applications.
    \newblock {\em Nanotechnology}, 30, 01 2019.
    
    \bibitem{Comm.Pres.2.10.2019.Savinov}
    V~Savinov, N~Papasimakis, D~P Tsai, and N~I Zheludev.
    \newblock Optical anapoles.
    \newblock {\em Communications Physics}, 2:10, 2019.
    
    \bibitem{Nanophotonics.7.2017.Talebi}
    Nahid Talebi, Surong Guo, and Peter Aken.
    \newblock Theory and applications of toroidal moments in electrodynamics: Their
    emergence, characteristics, and technological relevance.
    \newblock {\em Nanophotonics}, 7, 1 2017.
    
    \bibitem{LasPhotRev.13.1800266.2019.Gurvitz}
    Egor Gurvitz, Konstantin Ladutenko, Pavel Dergachev, Andrey Evlyukhin, Andrey
    Miroshnichenko, and Alexander Shalin.
    \newblock The high‐order toroidal moments and anapole states in
    all‐dielectric photonics.
    \newblock {\em Laser and Photonics Reviews}, 13:1800266, 2019.
    
    \bibitem{ScRep.5.2016.Zagoskin}
    Alexandre Zagoskin, A.~Chipouline, Evgeni Il'ichev, Robert Johansson, and
    Franco Nori.
    \newblock Toroidal qubits: Naturally-decoupled quiet artificial atoms.
    \newblock {\em Scientific Reports}, 5, 2014.
    
    \bibitem{PhysLettB.722.341.2013.Ho}
    C.~M. Ho and R.~J. Scherrer.
    \newblock Anapole dark matter.
    \newblock {\em Phys. Lett. B}, 722:341, 2013.
    
    \bibitem{NuclPhysB.907.1.2016.Cabral}
    Luis~G. Cabral-Rosetti, Myriam Mondrag{\'{o}}n, and Esteban Reyes-P{\'{e}}rez.
    \newblock {Anapole moment of the lightest neutralino in the cMSSM}.
    \newblock {\em Nucl. Phys. B}, 907:1, 2016.
    
    \bibitem{PhysRevD.32.1266.Radescu}
    E.~E. Radescu.
    \newblock On the electromagnetic properties of majorana fermions.
    \newblock {\em Phys. Rev. D}, 32:1266, 1985.
    
    \bibitem{ModPhysA.13.5257.1998.Dubovik}
    Dubovik~Vladimir M. and Valentin~E. Kuznetsov.
    \newblock The toroid dipole moment of the neutrino.
    \newblock {\em International Journal of Modern Physics A}, 13:5257, 1998.
    
    \bibitem{AnnPhys.209.13.1991.Costescu}
    A~Costescu and E.E Radescu.
    \newblock Dynamic toroid polarizability of atomic hydrogen.
    \newblock {\em Ann. Phys.}, 209:13, 1991.
    
    \bibitem{JPA30.3515.1997.Anghel}
    D.~V. Anghel.
    \newblock Mathematical considerations regarding the toroidal momentum operator.
    \newblock {\em J. Phys. A: Math. Gen.}, 30:3515--3525, 1997.
    
    \bibitem{PhysRep.187.145.1990.Dubovik}
    V.M. Dubovik and V.V. Tugushev.
    \newblock Toroid moments in electrodynamics and solid-state physics.
    \newblock {\em Phys. Rep.}, 187:145, 1990.
    
    \bibitem{Phys.Rev.C.65.1.2002.Haxton}
    W~C Haxton.
    \newblock {Nuclear anapole moments}.
    \newblock {\em Phys. Rev. C}, 65:1, 2002.
    
    \bibitem{NJP.9.324.2007.Marinov}
    K~Marinov, A~D Boardman, V~A Fedotov, and N~Zheludev.
    \newblock Toroidal metamaterial.
    \newblock {\em New Journal of Physics}, 9:324, 2007.
    
\end{thebibliography}
